\documentclass[a4paper,11pt]{article}
\pdfoutput=1 
\usepackage{jheppub} %%%%
\usepackage{array}
\usepackage{multirow}
\usepackage{amsmath}
\usepackage{slashed}
\usepackage{amstext,amssymb}
\usepackage{graphicx,wrapfig,lipsum}
\usepackage{xmpmulti}
\usepackage{animate}
\usepackage{epstopdf}
\usepackage{xcolor,colortbl}
\usepackage{multimedia}
\usepackage{url}
\usepackage{xspace}
\usepackage{dsfont}
\usepackage{subcaption}
\usepackage{placeins}
\usepackage{tabularx}
\usepackage{booktabs}
\usepackage{bm}
\usepackage{orcidlink}
\usepackage{color}
\usepackage{units}
\usepackage{braket}
\usepackage{makecell}
\usepackage{bigstrut}
\usepackage{tikz}
\usetikzlibrary{snakes}

\usepackage{mathtools}
\usepackage[T1]{fontenc} % if needed
\title{Probing Non-Holomorphic Modular $A_4$ Double Seesaw: Signatures in Neutrino Oscillation Experiments and Implications for Leptogenesis}

\author[a]{Pratik Adarsh\orcidlink{0009-0006-5526-4830},}
\author[b]{Dinesh Kumar Singha\orcidlink{0000-0001-7456-4691},}
\author[c]{Mitesh Kumar Behera\orcidlink{0000-0002-2137-3100},}
\author[a]{Sudhanwa Patra\orcidlink{0000-0002-7469-2279}}

\affiliation[a]{Department of Physics, Indian Institute of Technology Bhilai, Kutelabhata, Durg 491002, India}
\affiliation[b]{Center of Excellence for Advanced Materials and Sensing Devices, Ruđer Bošković Institute, 10000 Zagreb, Croatia}
\affiliation[c]{Department of Physics, School of Advanced Sciences, Vellore Institute of Technology, Tiruvalam Rd, Katpadi, Vellore, Tamil Nadu 632014, India.}
\emailAdd{pratikad@iitbhilai.ac.in}
\emailAdd{disingha@irb.hr}
\emailAdd{mitesh.behera@vit.ac.in}
\emailAdd{sudhanwa@iitbhilai.ac.in}

\abstract{We realize the double seesaw mechanism within a non-holomorphic modular $A_4$ framework by extending the Standard Model with three generations of right handed neutrinos (RHNs), three left-handed sterile neutrino fields, and an $A_4$-singlet scalar. The modular construction simultaneously forbids a bare Majorana mass term for the RHNs and, within the allowed parameter space, realizes the hierarchy required for the double seesaw, such that the RHN masses are induced through the heavier sterile neutrino sector. In this framework, light active neutrino masses, oscillation phenomenology, and the baryon asymmetry of the Universe emerge as interconnected consequences of the same underlying construction. A comprehensive scan of the modular parameter space yields viable solutions for normal ordering that reproduce the neutrino oscillation observables within the allowed ranges of global fit. We further examine the testability of the allowed parameter space against the projected sensitivities of DUNE, T2HK, and JUNO. DUNE and T2HK provide strong sensitivity to the atmospheric mixing parameters and significantly constrain the allowed model parameter space, whereas JUNO offers complementary high-precision probes of the solar mixing angle and the solar mass-squared splitting. The oscillation constrained parameter points simultaneously determine the induced RHN mass spectrum and the corresponding complex Yukawa textures, providing the ingredients required for thermal leptogenesis. For a representative unflavored benchmark in the strong-washout regime, the numerical solution of the Boltzmann equations including decays and inverse decays yields $Y_{\Delta B}\simeq8.11\times10^{-11}$, close to the observed baryon asymmetry. We also find that the oscillation-compatible parameter space admits an $N_1$-dominated two flavor thermal leptogenesis realization within a hierarchical, non-resonant RHN spectrum. Our results establish the non-holomorphic modular $A_4$ double seesaw framework as a predictive setting that links low-energy neutrino phenomenology with successful thermal leptogenesis, with its oscillation predictions directly testable at forthcoming precision neutrino experiments.}


\begin{document} 
\maketitle
%\flushbottom
%%%%%%%%%%%%%%%%%%%%%%%%%%%%%%%
\section{Introduction}\label{sec:intro}

The seesaw mechanism, conventionally realized through the Type-I \cite{Minkowski:1977sc,Mohapatra:1979ia,Yanagida:1979as,Gell-Mann:1979vob}, Type-II \cite{Magg:1980ut,Schechter:1980gr,Cheng:1980qt,Lazarides:1980nt,Mohapatra:1980yp}, and Type-III \cite{Foot:1988aq,Ma:2001kg,Ma:2002pf,Barr:2005je,Dorsner:2006fx,He:2012ub} scenarios, remains at the heart of physics beyond the Standard Model (BSM), addressing several shortcomings of the Standard Model (SM). In particular, it provides a natural explanation for the origin of tiny neutrino masses, as necessitated by neutrino oscillation experiments \cite{SNO:2002tuh,Super-Kamiokande:1998kpq,DayaBay:2012fng,DoubleChooz:2011ymz}. Among its other implications, the seesaw mechanism can also accommodate the observed neutrino mixing pattern. It further provides the necessary ingredients for realizing leptogenesis \cite{Fukugita:1986hr,Davidson:2008bu}, offering a compelling explanation for the observed baryon asymmetry of the Universe (BAU) \cite{Planck:2018vyg}, which remains unexplained within the SM.

Leptogenesis refers to a class of scenarios in which the observed BAU is explained through the dynamical generation of a lepton asymmetry before the electroweak phase transition (EWPT), which is partially converted into a baryon asymmetry through the $B+L$ violating SM sphaleron processes \cite{Klinkhamer:1984di,Arnold:1987mh,Kuzmin:1985mm,Rubakov:1996vz}. The realization of leptogenesis is model-dependent, and among the well-motivated scenarios, the Type-I seesaw provides the simplest and most widely studied framework \cite{Fukugita:1986hr,Davidson:2008bu}. By extending the SM particle content with gauge-singlet right-handed neutrinos (RHNs), it provides the essential ingredients required for leptogenesis. The Majorana nature of the RHNs leads to lepton number violation (LNV), while their complex Yukawa couplings with the SM Higgs and lepton doublets provide the required CP violation. The out of equilibrium dynamics of RHN decays in the expanding Universe then provide the departure from thermal equilibrium necessary for generating the lepton asymmetry.

Despite its simplicity and phenomenological success, minimal Type-I leptogenesis leaves several fundamental questions unanswered. In particular, the Majorana masses of the RHNs are introduced as independent input parameters, while their dynamical origin remains unspecified. For thermal leptogenesis at temperatures where the charged-lepton Yukawa interactions are out of equilibrium, the generated lepton state can be treated coherent, leading to the unflavored or single-flavor regime \cite{Abada:2006fw,Nardi:2006fx,Abada:2006ea,Blanchet:2006be,Davidson:2008bu}. In minimal Type-I unflavored thermal leptogenesis, there is generally no direct connection between the generated BAU and the low-energy neutrino oscillation parameters, since the CP asymmetry depends on high-energy parameters that are not fixed by the low-energy neutrino sector \cite{Branco:2001pq,Davidson:2008bu,Fong:2012buy}. Consequently, without imposing additional assumptions or flavor structures, the mechanism remains difficult to probe indirectly through low-energy neutrino observables.

These theoretical and phenomenological limitations motivate consideration of extended seesaw frameworks that involve additional neutral fermions and multiple mass scales. Various studies of leptogenesis have been carried out by realizing extended seesaw mechanisms within different BSM frameworks \cite{Gu:2010xc,Blanchet:2010kw,Agashe:2018cuf}. Among them, the double seesaw (DSS) mechanism \cite{Mohapatra:1986bd,Gu:2010xc, Patel:2023voj} provides a particularly intriguing framework in which an effective Majorana mass matrix for the RHNs is induced through additional sterile singlet fermions carrying large Majorana masses ($M_{S_i}$). In its characteristic realization, the neutral-fermion sector follows the hierarchy $M_S \gg M_{RS} \gg M_D$, where $M_{RS}$ denotes the RHN-sterile neutrino Dirac mass scale, while a bare Majorana mass term for the RHNs is absent. After integrating out the heavier sterile states, the RHNs acquire an effective Majorana mass through the underlying seesaw structure and can subsequently undergo CP-violating decays, leading to a lepton asymmetry analogous to the Type-I leptogenesis scenario. In suitably constrained realizations, the DSS framework can further establish correlations between the high-energy parameters governing leptogenesis and the low-energy neutrino oscillation observables \cite{Patel:2023voj}.

The leptogenesis dynamics, however, can depend significantly on the underlying framework in which the DSS mechanism is realized. For example, in Ref.~\cite{Patel:2023voj}, the DSS mechanism is embedded within a left-right symmetric model (LRSM) \cite{Mohapatra:1974gc,Pati:1974yy}. In this realization, the otherwise Type-I-like leptogenesis dynamics are modified by additional right-handed gauge interactions mediated by $W_R$ and $Z'$. Consequently, the viability of leptogenesis depends additionally on the masses and couplings of these gauge bosons, introducing further model dependence beyond the conventional Type-I-like scenario \cite{Frere:2008ct,Patel:2023voj}. Table~\ref{tab:DSSRealization} summarizes representative model frameworks realizing the DSS mechanism and their corresponding implications for leptogenesis.
\begin{table}[t]
	\centering
	\scriptsize
	\renewcommand{\arraystretch}{1.25}
	\begin{tabularx}{\textwidth}{
			>{\raggedright\arraybackslash}p{0.22\textwidth}
			>{\raggedright\arraybackslash}X
			>{\raggedright\arraybackslash}X
			>{\raggedright\arraybackslash}X
		}
		\toprule
		\textbf{Framework} & \textbf{DSS realization} & \textbf{Salient feature} & \textbf{Leptogenesis aspect} \\
		\midrule
		LRSM with scalar doublets \cite{Gu:2010xc,Patel:2023voj}
		&
		$N_R$ is part of the RH lepton doublet; adding $S_L$ realizes the DSS structure.
		&
		Gauge-motivated origin of $N_R$ and bare $N_R$ Majorana mass is forbidden.
		&
		Extra $W_R$ and $Z'$ interactions introduce additional scattering and washout effects.
		\\
		\midrule
		Gauged $U(1)_{B-L}$ extension \cite{Dey:2021ecr}
		&
		$N_R$ is motivated by anomaly cancellation; with suitable charges, bare $N_RN_R$ is forbidden, $S_LS_L$ is allowed, and $N_RS_L$ is generated after $B-L$ breaking.
		&
		The absence of a bare RHN Majorana mass can follow from gauge charges.
		&
		The extra $Z'$ and $B-L$ breaking scalar can introduce additional interactions affecting RHN production and washout.
		\\
		\midrule
		SO(10) / GUT embeddings \cite{Smirnov:2018luj}
		&
		RH neutrinos are embedded in the SO(10) matter multiplets, while additional neutral fermions generate the DSS structure after GUT symmetry breaking.
		&
		Connects DSS with quark-lepton unification.
		&
		GUT-scale fields and threshold effects can make leptogenesis more model dependent.
		\\
		\midrule
		Conventional $A_4$ flavor DSS \cite{Nanda:2025fvw}
		&
		$A_4$ symmetry controls the textures of $M_D$, $M_{RS}$ and $M_S$ through flavon alignments.
		&
		Predictive double seesaw flavor structure.
		&
		Leptogenesis is Type-I-like, but its flavor structure depends on flavon vacuum alignment.
		\\
		\midrule
		\textbf{This work:} non-holomorphic modular $A_4$ DSS
		&
		SM is extended by $3N_R$ and $3S_L$ fields + one $A_4$-singlet scalar $\rho$; the DSS structure follows from the modular-weight selection rules.
		&
		Flavor structure is constrained by $\tau$ without extra gauge bosons.
		&
		Leptogenesis retains Type-I-like thermal dynamics without additional gauge-mediated washout processes.
		\\
		\bottomrule
	\end{tabularx}
\caption{Representative realizations of the double seesaw mechanism in different BSM frameworks and their salient implications for leptogenesis. The non-holomorphic modular $A_4$ realization considered in this work is included for comparison.}
	\label{tab:DSSRealization}
\end{table}

Motivated by these considerations, in the present work, we realize the DSS mechanism within a non-holomorphic modular $A_4$ framework \cite{Qu:2024rns} and investigate the resulting leptogenesis phenomenology. The finite modular group $\Gamma_3 \simeq A_4$ provides a modular realization of the non-Abelian discrete $A_4$ flavor symmetry \cite{Ma:2001dn,Feruglio:2017spp}, which has been widely employed to describe the observed patterns of lepton masses and mixing \cite{Loualidi:2025tgw,Behera:2026tmo,Behera:2020sfe,Mishra:2022egy,Altarelli:2005yp,Altarelli:2005yx,Ishimori:2010au,King:2013eh}. In conventional discrete flavor constructions, the flavor symmetry is spontaneously broken by scalar fields known as flavons, whose vacuum expectation values are required to follow specific directions in flavor space \cite{Altarelli:2005yp,Altarelli:2005yx,Ishimori:2010au,King:2013eh}. Such constructions can involve multiple flavon fields, nontrivial vacuum-alignment mechanisms, and auxiliary symmetries introduced to forbid unwanted operators, thereby increasing model-building complexity and reducing its predictive power \cite{Ishimori:2010au,King:2013eh}.

Modular flavor symmetry offers an appealing alternative in which the Yukawa couplings transform as modular forms under a finite modular group $\Gamma_N$ \cite{Feruglio:2017spp,Ding:2023htn}. Such symmetries are well motivated by compactified extra-dimensional and string constructions in which modular invariance arises naturally \cite{Feruglio:2017spp,Feruglio:2019ybq,Ohki:2020bpo}. In minimal modular constructions, flavor-symmetry breaking is governed primarily by the vacuum expectation value of the complex modulus $\tau \equiv x+i\,y$, thereby avoiding the need for flavons transforming nontrivially under the flavor symmetry and the associated vacuum-alignment problem \cite{Feruglio:2017spp}. Early modular flavor models were primarily formulated in supersymmetric frameworks, where the holomorphicity of the superpotential requires the Yukawa couplings to be holomorphic modular forms, thereby restricting the admissible modular weights and flavor structures. The absence of experimental evidence for low-energy supersymmetry (SUSY) has spurred growing interest in non-holomorphic modular flavor symmetries \cite{Qu:2024rns}. In such constructions, the Yukawa couplings are not subject to the holomorphicity condition imposed by a superpotential and can instead be described in terms of polyharmonic Maa{\ss} forms satisfying the corresponding Laplacian condition \cite{Qu:2024rns}. These forms can carry positive, zero, and negative modular weights, allowing a wider range of flavor structures while retaining a constrained parameter space \cite{Qu:2024rns,Ding:2024inn}.

Realizing the DSS mechanism within a non-holomorphic modular $A_4$ framework offers a comparatively minimal and predictive setup. In the present construction, the modular-weight assignments forbid a bare Majorana mass term for the RHNs, such that their Majorana masses arise entirely through the underlying double seesaw structure. In addition to the three RHNs and three sterile neutrino fields required for the DSS mechanism, the model contains a scalar flavon $\rho$ assigned to the trivial singlet representation of $A_4$. Consequently, its vacuum expectation value does not require a nontrivial alignment in flavor space. Moreover, the model does not involve an enlarged gauge sector or flavons transforming in nontrivial $A_4$ representations. After the heavier sterile neutrino states are integrated out, leptogenesis proceeds through the CP-violating decays of the effective Majorana RHNs via their Yukawa interactions, without additional gauge-mediated scattering and washout processes \cite{Fukugita:1986hr,Gu:2010xc,Davidson:2008bu}.

The modular $A_4$ symmetry constrains the Yukawa textures and correlates the high-energy parameters relevant for leptogenesis with low-energy neutrino observables. We perform a numerical scan over the modular parameter space and confront the resulting neutrino masses and mixing parameters with the current global-fit data from \textsf{NuFIT} \cite{Esteban:2024eli}. The surviving parameter space is further compared with the projected sensitivities of DUNE, T2HK, and JUNO \cite{DUNE:2020ypp,Hyper-Kamiokande:2018ofw,JUNO:2021vlw} to assess the future testability of the model. Each viable parameter set determines correlated values of the lightest neutrino mass, the Dirac neutrino mass matrix, the induced RHN mass spectrum, and the CP-violating parameters entering the leptogenesis analysis. Using representative benchmark points from the allowed parameter space, we demonstrate that the observed BAU \cite{Planck:2018vyg} can be successfully reproduced. The present framework therefore combines high-scale Type-I-like leptogenesis with a dynamical origin of the RHN Majorana masses and testable connections to low-energy neutrino phenomenology.

The paper is structured as follows. In Section~\ref{non-holomorphic}, we introduce the non-holomorphic modular $A_4$ double seesaw framework, present the particle content and modular assignments, construct the lepton mass matrices, and derive the resulting light neutrino and effective RHN mass matrices. In Section~\ref{sec:Exp_th_details}, we discuss the neutrino oscillation observables, the simulation details of the neutrino oscillation experiments, and the numerical procedure used to scan the model parameter space and obtain the oscillation predictions. In Section~\ref{sec:results}, we present the model predictions, the allowed parameter space, and its sensitivity to next-generation neutrino oscillation experiments. In Section~\ref{sec:leptogenesis}, we explore the unflavored and $N_1$-dominated two-flavor thermal leptogenesis scenarios allowed by the model. Finally, we summarize our findings in Section~\ref{sec:conclusion}.

\section{Non-holomorphic modular $A_4$ double seesaw framework}
\label{non-holomorphic}

\subsection{Non-holomorphic modular symmetry}
The modular-symmetry approach attributes the observed fermion flavor structure to the transformation properties of matter fields and Yukawa couplings under the modular group. In contrast to conventional discrete flavor models, where spontaneous symmetry breaking is typically implemented through flavon fields, modular-invariant constructions encode the flavor dependence directly in modular forms of the complex modulus $\tau$. In this work, we adopt the non-supersymmetric formulation of non-holomorphic modular flavor symmetry based on polyharmonic Maa{\ss} forms introduced in Ref.~\cite{Qu:2024rns}.
The complex modulus $\tau$, defined in the upper half-plane with $\mathrm{Im}\,\tau>0$, transforms under the action of $\gamma\in SL(2,\mathbb{Z})$ through the Möbius transformation
\begin{equation}
	\tau \rightarrow \gamma\tau=
	\frac{a\tau+b}{c\tau+d},
	\qquad
	\gamma=
	\begin{pmatrix}
		a & b\\
		c & d
	\end{pmatrix}
	\in SL(2,\mathbb{Z}),
\end{equation}
where $a$, $b$, $c$, and $d$ are integers satisfying the unimodularity condition $ad-bc=1$. Since $\gamma$ and $-\gamma$ induce the same transformation of $\tau$, the modular group can be identified as $\overline{\Gamma}\equiv PSL(2,\mathbb{Z})$. Introducing the principal congruence subgroup $\overline{\Gamma}(N)$ of level $N$ defines the finite modular group $\Gamma_N\equiv\overline{\Gamma}/\overline{\Gamma}(N)$, which has been extensively employed in constructing predictive flavor models
\cite{Feruglio:2017spp,Ding:2023htn,Kobayashi:2019gtp,Kashav:2021zir,Gogoi:2023jzl,Pathak:2024sei,Behera:2022wco}.
Within this framework, the matter multiplets $\psi$ and $\psi^c$, the Higgs field $H$, and the polyharmonic Maa{\ss} forms $Y_r^{(k_Y)}(\tau)$ are assigned modular weights $k_\psi$, $k_{\psi^c}$, $k_H$, and $k_Y$, respectively. They transform according to the irreducible representations $\rho_\psi$, $\rho_{\psi^c}$, $\rho_H$, and $\rho_Y$ of the finite modular group. Under a modular transformation, a matter multiplet $\psi$ and a modular form $Y_r^{(k_Y)}(\tau)$ transform as $\psi\rightarrow(c\tau+d)^{-k_\psi}\rho_\psi(\gamma)\psi$ and $Y_r^{(k_Y)}(\tau)\rightarrow(c\tau+d)^{k_Y}\rho_Y(\gamma)Y_r^{(k_Y)}(\tau)$, respectively, with analogous transformations for $\psi^c$ and $H$. The modular invariance of a Yukawa interaction involving $Y_r^{(k_Y)}\psi\psi^cH$ requires the total modular weight to vanish and the tensor product of the corresponding representations to contain the trivial singlet \cite{Qu:2024rns}. These conditions are expressed as
\begin{equation}
	k_Y = k_\psi + k_{\psi^c} + k_H,
	\qquad
	\rho_Y \otimes \rho_\psi \otimes \rho_{\psi^c} \otimes \rho_H
	\supset \mathbf{1}.
\end{equation}
Consequently, once the modular weights and representation assignments of the fields are specified, modular invariance restricts the admissible Yukawa operators and their associated flavor structures.

\subsection{Model setup and field assignments}
\label{sec:model_FW}

We now specify the particle content and symmetry assignments employed in the present construction. The SM gauge sector remains unchanged, while the neutral-fermion sector is extended by three right-handed neutrinos, $N_R=(N_{R_1},N_{R_2},N_{R_3})^T$, and three left-handed sterile neutrino fields, $S_L=(S_{L_1},S_{L_2},S_{L_3})^T$. We further introduce a gauge-singlet scalar flavon $\rho$, whose vacuum expectation value generates the Dirac mass term connecting $N_R$ and $S_L$. Both $N_R$ and $S_L$ transform as triplets under the modular $A_4$ symmetry, whereas the SM Higgs doublet $H$ and the scalar flavon $\rho$ are assigned to the trivial singlet representation. The three lepton doublets and the corresponding right-handed charged leptons\footnote{For the charged-lepton sector, we use the notation $(L_1,L_2,L_3)=(L_e,L_\mu,L_\tau)$ and $(E_1,E_2,E_3)=(e_R,\mu_R,\tau_R)$.} are assigned to the $A_4$ singlet representations $\mathbf{1}$, $\mathbf{1}^{\prime}$, and $\mathbf{1}^{\prime\prime}$, respectively. The complete field content and the corresponding gauge, $A_4$, and modular-weight assignments are summarized in Table~\ref{table:particle_content}.

\begin{table}[t]
	\centering
	\scriptsize
	\renewcommand{\arraystretch}{1.2}
	\begin{tabular}{c|ccc|ccc|cccc}
		\toprule
		Fields
		& $E_1$ & $E_2$ & $E_3$
		& $L_1$ & $L_2$ & $L_3$
		& $N_R$ & $S_L$ & $H$ & $\rho$ \\
		\midrule
		$SU(2)_L$
		& $\mathbf{1}$ & $\mathbf{1}$ & $\mathbf{1}$
		& $\mathbf{2}$ & $\mathbf{2}$ & $\mathbf{2}$
		& $\mathbf{1}$ & $\mathbf{1}$ & $\mathbf{2}$ & $\mathbf{1}$ \\
		
		$U(1)_Y$
		& $-1$ & $-1$ & $-1$
		& $-1/2$ & $-1/2$ & $-1/2$
		& $0$ & $0$ & $1/2$ & $0$ \\
		
		$A_4$
		& $\mathbf{1}$ & $\mathbf{1}^{\prime}$ & $\mathbf{1}^{\prime\prime}$
		& $\mathbf{1}$ & $\mathbf{1}^{\prime}$ & $\mathbf{1}^{\prime\prime}$
		& $\mathbf{3}$ & $\mathbf{3}$ & $\mathbf{1}$ & $\mathbf{1}$ \\
		
		$k_I$
		& $1/2$ & $1/2$ & $1/2$
		& $-1/2$ & $-1/2$ & $-1/2$
		& $-3/2$ & $0$ & $0$ & $-1/2$ \\
		\bottomrule
	\end{tabular}
	\caption{Particle content of the model and the corresponding transformation properties under $SU(2)_L\times U(1)_Y\times A_4$. The last row specifies the modular weight $k_I$ assigned to each field.}
	\label{table:particle_content}
	
\end{table}

The use of half-integral modular weights has been explored in several modular-flavor constructions \cite{Liu:2020msy,Zhang:2025dsa,Majhi:2026jdk}. Within the present non-holomorphic modular framework, the assignments listed in Table~\ref{table:particle_content} are chosen such that the interactions required for the DSS mechanism remain invariant under the imposed symmetries. Together with the available polyharmonic Maa{\ss} forms, these assignments determine the admissible lepton sector operators. In particular, they allow the terms responsible for the charged lepton masses, the active neutrino-RHN Dirac mass matrix, the RHN-sterile neutrino Dirac mass matrix, and the sterile neutrino Majorana mass matrix. At the same time, the modular-weight selection rules forbid a bare Majorana mass term for the RHNs, ensuring that their Majorana masses arise from the underlying double seesaw structure.

Although $\rho$ develops a nonzero vacuum expectation value, it transforms as a trivial singlet under $A_4$ and therefore does not require a nontrivial vacuum alignment in flavor space. Since both $N_R$ and $S_L$ transform as $A_4$ triplets, their tensor product allows singlet, symmetric-triplet, and antisymmetric-triplet contractions. Together with the modular-weight assignments, these contractions determine the allowed flavor structures of the lepton mass matrices constructed in the following subsection.

\subsection{Lepton mass terms and flavor structures}
\label{subsec:lepton_mass_terms}

Using the field assignments summarized in Table~\ref{table:particle_content}, we now construct the lepton-sector interactions allowed by the gauge and modular $A_4$ symmetries. The scalar fields acquire vacuum expectation values according to
\begin{equation}
	\langle H\rangle=
	\frac{1}{\sqrt{2}}
	\begin{pmatrix}
		0\\
		v_H
	\end{pmatrix},
	\qquad
	\langle\rho\rangle=\frac{v_\rho}{\sqrt{2}},
	\label{scalarVEV}
\end{equation}
where $v_H=246~\mathrm{GeV}$. These interactions generate the charged-lepton mass matrix $M_\ell$, the active neutrino-RHN Dirac mass matrix $M_D$, the RHN-sterile neutrino Dirac mass matrix $M_{RS}$, and the sterile neutrino Majorana mass matrix $M_S$.\\

\noindent\textbf{\underline{Charged-lepton sector:}}\\
The assignments of the lepton doublets and right-handed charged leptons allow the charged-lepton Yukawa interactions without the need to insert any nontrivial modular form. The relevant Lagrangian is
\begin{equation}
L_\ell = y_\ell^{e} \overline{L_1} H E_1 + y_\ell^\mu  \overline{L_2} H E_2 + y_\ell^\tau \overline{L_3} H E_3 + h.c. 
	\label{eq:charged_lepton_lagrangian}
\end{equation}
Since $L_{1,2,3}$ transform as $\mathbf{1}$, $\mathbf{1}^{\prime}$, and $\mathbf{1}^{\prime\prime}$, respectively, their Dirac conjugates transform as $\mathbf{1}$, $\mathbf{1}^{\prime\prime}$, and $\mathbf{1}^{\prime}$. Consequently, each term in Eq.~\eqref{eq:charged_lepton_lagrangian} forms a trivial $A_4$ singlet through
$\mathbf{1}\otimes\mathbf{1}=\mathbf{1}$,
$\mathbf{1}^{\prime\prime}\otimes\mathbf{1}^{\prime}=\mathbf{1}$, and
$\mathbf{1}^{\prime}\otimes\mathbf{1}^{\prime\prime}=\mathbf{1}$ \cite{Ma:2001dn}.
After electroweak symmetry breaking, the charged-lepton mass matrix is therefore diagonal and takes the form
\begin{equation}
	M_\ell
	=
	\frac{v_H}{\sqrt{2}}
	\begin{pmatrix}
		y_\ell^e & 0 & 0\\
		0 & y_\ell^\mu & 0\\
		0 & 0 & y_\ell^\tau
	\end{pmatrix}
	=
	\begin{pmatrix}
		m_e & 0 & 0\\
		0 & m_\mu & 0\\
		0 & 0 & m_\tau
	\end{pmatrix}.
	\label{eq:charged_lepton_mass_matrix}
\end{equation}
Here $m_e, m_\mu$ and $m_\tau$ are the observed charged-lepton masses. The analysis is thus performed in the basis where the charged-lepton mass matrix is diagonal, and the lepton mixing originates entirely from the neutral-fermion sector.\\

\noindent\textbf{\underline{Active neutrino-RHN Dirac sector:}}\\
The right-handed neutrinos are arranged into an $A_4$ triplet,
$N_R=(N_{R_1},N_{R_2},N_{R_3})^T$. Their coupling to the lepton doublets requires the weight-$(-2)$ triplet polyharmonic Maa{\ss} form
\begin{equation}
	Y_{\mathbf{3}}^{(-2)}(\tau)
	=
	\left(
	Y_{3,1}^{(-2)},
	Y_{3,2}^{(-2)},
	Y_{3,3}^{(-2)}
	\right)^T.
\end{equation}
The modular-invariant Dirac-neutrino interactions can then be written as
\begin{align}
\mathcal{L}_{M_D} = \alpha_D \overline{L_1} \tilde{H} \left(N_R Y_3^{(-2)}\right)_1 + \beta_D \overline{L_2}\tilde{H} \left(N_R Y_3^{(-2)}\right)_{1^\prime} + \gamma_D \overline{L_3} \tilde{H} \left(N_R Y_3^{(-2)}\right)_{1^{\prime\prime}} + h.c.,
	\label{eq:dirac_neutrino_lagrangian}
\end{align}
where $\widetilde{H}=i\sigma_2H^*$. Using the $A_4$ multiplication rules and defining $\widetilde{\beta}_D=\beta_D/\alpha_D$ and $\widetilde{\gamma}_D=\gamma_D/\alpha_D$, the corresponding Dirac mass matrix is obtained as
\begin{equation}
	M_D = \frac{v_H}{\sqrt{2}} \alpha_D
	\begin{pmatrix}
		1 & 0& 0 \\
		0 & \widetilde{\beta}_D & 0 \\
		0 & 0 & \widetilde{\gamma}_D
	\end{pmatrix}
	\begin{pmatrix}
		Y_{3,1}^{(-2)} &~~  Y_{3,3}^{(-2)} &~~  Y_{3,2}^{(-2)}\\
		Y_{3,2}^{(-2)} &~~  Y_{3,1}^{(-2)} &~~  Y_{3,3}^{(-2)}\\
		Y_{3,3}^{(-2)} &~~ Y_{3,2}^{(-2)} & ~~ Y_{3,1}^{(-2)}\\
	\end{pmatrix}.
	\label{eq:dirac_neutrino_mass_matrix}
\end{equation}
Thus, the flavor structure of $M_D$ is governed by the modulus $\tau$ through the components of $Y_{\mathbf{3}}^{(-2)}(\tau)$, together with the two independent coupling ratios $\widetilde{\beta}_D$ and $\widetilde{\gamma}_D$.\\

\noindent\textbf{\underline{RHN-sterile neutrino Dirac sector:}}\\
Both $N_R$ and $S_L$ transform as triplets under $A_4$. Their product decomposes as
\begin{equation}
	\mathbf{3}\otimes\mathbf{3}
	=
	\mathbf{1}
	\oplus\mathbf{1}^{\prime}
	\oplus\mathbf{1}^{\prime\prime}
	\oplus\mathbf{3}_S
	\oplus\mathbf{3}_A,
\end{equation}
allowing both symmetric and antisymmetric triplet contractions. The interactions responsible for the RHN-sterile neutrino Dirac mass matrix are
\begin{align}
	-\mathcal{L}_{M_{RS}}
	={}&
	\Big[
	\alpha_{RS}
	\left(
	\left(\overline{S_L}N_R\right)_{\mathbf{3}_S}
	Y_{\mathbf{3}}^{(-2)}
	\right)_{\mathbf{1}}
	+
	\beta_{RS}
	\left(
	\left(\overline{S_L}N_R\right)_{\mathbf{3}_A}
	Y_{\mathbf{3}}^{(-2)}
	\right)_{\mathbf{1}}
	\nonumber\\
	&\hspace{1.2cm}
	+
	\gamma_{RS}
	\left(\overline{S_L}N_R\right)_{\mathbf{1}}
	Y_{\mathbf{1}}^{(-2)}
	\Big]\rho
	+\mathrm{h.c.}
	\label{eq:RS_lagrangian}
\end{align}
Defining $\widetilde{\beta}_{RS}=\beta_{RS}/\alpha_{RS}$ and $\widetilde{\gamma}_{RS}=\gamma_{RS}/\alpha_{RS}$, the resulting mass matrix obtained after $\rho$ acquires its vacuum expectation value is
\begin{align}
	M_{RS}
	={}&
	\frac{v_\rho}{\sqrt{2}}\alpha_{RS}
	\Bigg[
	\begin{pmatrix}
		2Y_{3,1}^{(-2)}+\widetilde{\gamma}_{RS}Y_{\mathbf{1}}^{(-2)}
		&
		-Y_{3,3}^{(-2)}
		&
		-Y_{3,2}^{(-2)}
		\\
		-Y_{3,3}^{(-2)}
		&
		2Y_{3,2}^{(-2)}
		&
		-Y_{3,1}^{(-2)}+\widetilde{\gamma}_{RS}Y_{\mathbf{1}}^{(-2)}
		\\
		-Y_{3,2}^{(-2)}
		&
		-Y_{3,1}^{(-2)}+\widetilde{\gamma}_{RS}Y_{\mathbf{1}}^{(-2)}
		&
		2Y_{3,3}^{(-2)}
	\end{pmatrix}
	\nonumber\\
	&\hspace{1.0cm}
	+
	\widetilde{\beta}_{RS}
	\begin{pmatrix}
		0 & Y_{3,3}^{(-2)} & -Y_{3,2}^{(-2)}\\
		-Y_{3,3}^{(-2)} & 0 & Y_{3,1}^{(-2)}\\
		Y_{3,2}^{(-2)} & -Y_{3,1}^{(-2)} & 0
	\end{pmatrix}
	\Bigg].
	\label{eq:RS_mass_matrix}
\end{align}
The first matrix in Eq.~\eqref{eq:RS_mass_matrix} originates from the symmetric-triplet and singlet contractions, whereas the second matrix originates from the antisymmetric-triplet contraction. Consequently, $M_{RS}$ is generally not symmetric when $\beta_{RS}\neq0$.\\

\noindent\textbf{\underline{Sterile neutrino Majorana sector:}}\\
The modular assignments permit a bare Majorana mass term for $S_L$. Since the Majorana bilinear is symmetric in flavor space, only the symmetric-triplet and singlet contractions contribute. The corresponding Lagrangian is
\begin{equation}
	-\mathcal{L}_{M_S}
	=
	\frac{M_0}{2}
	\left[
	\alpha_S
	\left(
	Y_{\mathbf{3}}^{(0)}
	\left(\overline{S_L^c}S_L\right)_{\mathbf{3}_S}
	\right)_{\mathbf{1}}
	+
	\gamma_S
	Y_{\mathbf{1}}^{(0)}
	\left(\overline{S_L^c}S_L\right)_{\mathbf{1}}
	\right]
	+\mathrm{h.c.},
	\label{eq:sterile_majorana_lagrangian}
\end{equation}
where $M_0$ denotes the characteristic sterile neutrino mass scale. Using $Y_{\mathbf{1}}^{(0)}=1$ and defining $\widetilde{\gamma}_S=\gamma_S/\alpha_S$, the sterile neutrino Majorana mass matrix is obtained as
\begin{equation}
	M_S
	=
	M_0\alpha_S
	\left[
	\begin{pmatrix}
		2Y_{3,1}^{(0)} & -Y_{3,3}^{(0)} & -Y_{3,2}^{(0)}\\
		-Y_{3,3}^{(0)} & 2Y_{3,2}^{(0)} & -Y_{3,1}^{(0)}\\
		-Y_{3,2}^{(0)} & -Y_{3,1}^{(0)} & 2Y_{3,3}^{(0)}
	\end{pmatrix}
	+
	\widetilde{\gamma}_S
	\begin{pmatrix}
		1 & 0 & 0\\
		0 & 0 & 1\\
		0 & 1 & 0
	\end{pmatrix}
	\right].
	\label{eq:sterile_majorana_mass_matrix}
\end{equation}
Unlike the sterile neutrino Majorana term, a bare Majorana mass term for $N_R$ is forbidden by the modular-weight assignments. The RHN Majorana mass matrix is therefore generated effectively through the interplay of $M_{RS}$ and $M_S$, as discussed in the following subsection.

\subsection{Double seesaw realization and light-neutrino masses}
\label{subsec:double_seesaw}

The inclusion of three gauge-singlet sterile neutrinos $S_L$ allows the double seesaw mechanism to be realized within the present framework. Collecting the neutral-lepton mass terms obtained in the previous subsection, the mass Lagrangian can be written as
\begin{equation}
	-\mathcal{L}_{\mathrm{mass}}^\nu
	= \mathcal{L}_{M_D}+\mathcal{L}_{M_{RS}}+\mathcal{L}_{M_S}.
\end{equation}
After the scalar fields acquire VEVs ($\langle H\rangle$, $\langle \rho\rangle$) and thus lead to spontaneous symmetry breaking (SSB), the total $9\times9$ neutral fermion mass matrix in the flavor basis $\left(\nu_L, N_R^c, S_L \right)^T$ is given by
\begin{eqnarray}
	\mathbb{M}_\nu = \left(\begin{array}{c|ccc}   & \nu_L & N_R^c  & S_L   \\ \hline
		\nu_L  & 0       & M_D       & 0 \\
		N_R^c    & M^T_D         & 0       & M_{RS} \\
		S_L & 0     & M_{RS}^T    & M_S
	\end{array}
	\right).
	\label{eq:numatrix-complete}
\end{eqnarray}
With the assumed mass hierarchy $|M_D|\ll|M_{RS}|<|M_{S}|$, the mass matrix $\mathbb{M}_\nu$ can be approximately block diagonalized in two successive steps to give us the mass matrices of light and heavy neutrinos as follows \cite{Patra:2023ltl}:

\begin{eqnarray}
	m_{\nu} &=& M_D (M^{T}_{RS})^{-1} M_S M^{-1}_{RS}M^{T}_D, \\ \nonumber
	M_N &=& -M_{RS} M^{-1}_S M^{T}_{RS}, \\ \nonumber
	m_S &=& M_S. \\ \nonumber
	\label{eq:HLN}
\end{eqnarray}
For the numerical analysis, it is convenient to separate the overall mass scales from the dimensionless flavor structures. We write
\begin{equation}
	M_D
	=
	\frac{v_H}{\sqrt{2}}\alpha_D\widetilde{M}_D,
	\qquad
	M_{RS}
	=
	\frac{v_\rho}{\sqrt{2}}\alpha_{RS}\widetilde{M}_{RS},
	\qquad
	M_S
	=
	M_0\alpha_S\widetilde{M}_S,
	\label{eq:mass_matrix_rescaling}
\end{equation}
where the matrices $\widetilde{M}_D$, $\widetilde{M}_{RS}$, and $\widetilde{M}_S$ contain the dependence on the modulus $\tau$ and the dimensionless coupling ratios introduced in the previous subsection. The light-neutrino mass matrix can then be expressed as
\begin{equation}
	m_\nu
	=
	\kappa_\nu\widetilde{m}_\nu,
	\label{eq:scaled_neutrino_mass}
\end{equation}
where
\begin{equation}
	\kappa_\nu
	\equiv
	\left(\frac{v_H^2}{v_\rho^2}\right)
	\left(M_0\alpha_S\right)
	\left(\frac{\alpha_D^2}{\alpha_{RS}^2}\right),
	\label{eq:kappa1}
\end{equation}
and
\begin{equation}
	\widetilde{m}_\nu
	=
	\widetilde{M}_D
	\left(\widetilde{M}_{RS}^T\right)^{-1}
	\widetilde{M}_S
	\widetilde{M}_{RS}^{-1}
	\widetilde{M}_D^T.
	\label{eq:dimensionless_neutrino_mass}
\end{equation}
An overall phase contained in $\kappa_\nu$ can be absorbed, and its magnitude determines the absolute scale of the light-neutrino masses.

Since $\widetilde{m}_\nu$ is a complex symmetric matrix, it can be diagonalized through a Takagi decomposition,
\begin{equation}
	U_\nu^T\widetilde{m}_\nu U_\nu
	=
	\operatorname{diag}
	\left(
	\widetilde{m}_1,
	\widetilde{m}_2,
	\widetilde{m}_3
	\right),
	\label{eq:Takagi_neutrino}
\end{equation}
where $\widetilde{m}_i$ are real and non-negative. Equivalently, the mixing matrix can be obtained from
\begin{equation}
	U_\nu^\dagger
	\left(
	\widetilde{m}_\nu^\dagger\widetilde{m}_\nu
	\right)
	U_\nu
	=
	\operatorname{diag}
	\left(
	\widetilde{m}_1^2,
	\widetilde{m}_2^2,
	\widetilde{m}_3^2
	\right).
\end{equation}
The physical light-neutrino masses are then given by
\begin{equation}
	m_i
	=
	\left|\kappa_\nu\right|\widetilde{m}_i.
\end{equation}
Since the charged-lepton mass matrix is diagonal in the chosen basis, the lepton mixing matrix is directly determined by $U_\nu$, up to unphysical charged-lepton rephasings.

The overall scale $\left|\kappa_\nu\right|$ is fixed by the atmospheric mass-squared splitting. For normal ordering and inverted ordering, respectively, one obtains
\begin{align}
	\mathrm{NO}:\qquad
	\left|\kappa_\nu\right|^2
	&=
	\frac{\left|\Delta m_{\mathrm{atm}}^2\right|}
	{\widetilde{m}_3^2-\widetilde{m}_1^2},
	\nonumber\\
	\mathrm{IO}:\qquad
	\left|\kappa_\nu\right|^2
	&=
	\frac{\left|\Delta m_{\mathrm{atm}}^2\right|}
	{\widetilde{m}_2^2-\widetilde{m}_3^2}.
	\label{eq:kappa2}
\end{align}
The corresponding solar mass-squared splitting is
\begin{equation}
	\Delta m_{\mathrm{sol}}^2
	=
	\left|\kappa_\nu\right|^2
	\left(
	\widetilde{m}_2^2-\widetilde{m}_1^2
	\right),
	\label{eq:solar_mass_splitting}
\end{equation}
which is compared with its experimentally allowed range in the numerical analysis.

The effective RHN Majorana mass matrix can similarly be written as
\begin{equation}
	M_N
	=
	-\frac{v_\rho^2\alpha_{RS}^2}
	{2M_0\alpha_S}
	\widetilde{M}_N,
	\label{eq:scaled_RHN_mass}
\end{equation}
where
\begin{equation}
	\widetilde{M}_N
	=
	\widetilde{M}_{RS}
	\widetilde{M}_S^{-1}
	\widetilde{M}_{RS}^T.
\end{equation}
Using Eq.~\eqref{eq:kappa1}, its overall scale may also be expressed as
\begin{equation}
	M_N
	=
	-\frac{v_H^2\alpha_D^2}
	{2\kappa_\nu}
	\widetilde{M}_N.
	\label{eq:RHN_mass_kappa_relation}
\end{equation}
Equation~\eqref{eq:kappa1} shows that the low-energy neutrino data constrain only the combination of high-energy parameters entering $\kappa_\nu$, rather than fixing $v_\rho$, $M_0$, $\alpha_S$, $\alpha_D$, and $\alpha_{RS}$ individually. Consequently, for a fixed modular flavor structure and a value of $\kappa_\nu$ determined from the neutrino mass-squared differences, the overall RHN mass scale can be adjusted by varying $\alpha_D$ together with one or more of the remaining high-energy parameters such that $\kappa_\nu$ remains unchanged. This freedom is restricted by the perturbativity of the Yukawa couplings and by the mass hierarchy required for the validity of the double seesaw approximation\footnote{To ensure the hierarchy required for the double seesaw realization and for consistently integrating out the sterile neutrino states, we impose $\min_k m_{S_k}\simeq10^{2}\,\max_j m_{N_j}$ in the numerical analysis, consistent with the hierarchical setup of Ref.~\cite{Patra:2023ltl}.}.
%%%%%%%%%%%%%%%%%%%%%%%%%%%%%%%%%%%%%%%%%%%%%%%%%%%%%%%%%
\section{Neutrino oscillation observables and simulation setup}
\label{sec:Exp_th_details}

\subsection{Neutrino oscillation observables}
\label{Num_ana}

Within the standard three-neutrino framework, neutrino oscillations are described by six physical parameters: two independent mass-squared differences, $\Delta m^2_{\rm sol}$ and $\Delta m^2_{\rm atm}$, three mixing angles $\theta_{12}$, $\theta_{23}$, and $\theta_{13}$, and the Dirac CP phase $\delta_{\rm CP}$. The solar mass-squared splitting is defined as $\Delta m^2_{\rm sol} = m_2^2 - m_1^2$, independent of the neutrino mass ordering. The atmospheric mass-squared difference, by contrast, depends on the ordering: we define $\Delta m^2_{\rm atm}=m_3^2-m_1^2$ for Normal Ordering (NO) and $\Delta m^2_{\rm atm}=m_2^2-m_3^2$ for Inverted Ordering (IO).

The mixing angles are obtained from the elements of the PMNS matrix $U$ as
\begin{align}
	\sin^2\theta_{12} &= \frac{|U_{e2}|^2}{1 - |U_{e3}|^2}, \\
	\sin^2\theta_{23} &= \frac{|U_{\mu 3}|^2}{1 - |U_{e3}|^2}, \\
	\sin^2\theta_{13} &= |U_{e3}|^2 ,
\end{align}
while, in the standard PMNS parametrization \cite{ParticleDataGroup:2018ovx}, the Dirac CP phase is encoded in the complex phase structure of $U_{e3}$.

We determine the oscillation-compatible parameter space by constraining our model inputs with the neutrino oscillation observables from global analyses~\cite{deSalas:2020pgw,Capozzi:2021fjo,Esteban:2024eli}. In the numerical analysis, the measured oscillation parameters are required to lie within their corresponding $3\sigma$ ranges, with the exception of the Dirac CP phase $\delta_{\rm CP}$. Since its present experimental determination remains comparatively weak, we allow $\delta_{\rm CP}$ to vary over its entire physical range of $[-180^\circ,180^\circ]$ and treat it as a prediction of the model. In our parameter space scan, we utilized benchmark points corresponding to both the Lower Octant (LO) and Higher Octant (HO). Because the global $3\sigma$ allowed region is continuous across both octants, our model's predictions successfully span this entire $3\sigma$ interval, inherently accommodating both LO and HO solutions regardless of the initial benchmark. The best-fit values and $3\sigma$ intervals from \textsf{NuFIT}~6.0 used in this study are summarized in Table~\ref{table:data_nufit}. The quoted interval for $\delta_{\rm CP}$ is shown for comparison and is not imposed as a constraint in the scan.

\begin{table}[t]
	\centering
	\begin{tabular}{|l|c|c|}
		\hline\hline
		\multirow{2}{*}{Parameters} & \multicolumn{2}{c|}{Normal ordering} \\ \cline{2-3}
		&Best-fit value & $3\sigma$ range \\ \hline
		$\sin^2\theta_{12}$ & $0.308_{-0.011}^{+0.012}$ & $[0.275, 0.345]$ \\
		$\sin^2\theta_{23}$ (LO)& $0.470_{-0.013}^{+0.017}$ & $[0.435, 0.585]$ \\
		$\sin^2\theta_{23}$ (HO)& $0.550_{-0.015}^{+0.012}$ & $[0.440, 0.584]$ \\
		$\sin^2\theta_{13}$ & $0.02215_{-0.00058}^{+0.00056}$ & $[0.02030, 0.02388]$ \\
		\hline
		$\frac{\Delta m_{\mathrm{sol}}^2}{10^{-5} \mathrm{eV^2}}$ & $7.49_{-0.19}^{+0.19}$ & $[6.92, 8.05]$ \\
		$\frac{\Delta m_{\mathrm{atm}}^2}{10^{-3} \mathrm{eV^2}}$
		& $2.513_{-0.019}^{+0.021}$ & $[2.451, 2.578]$ \\
		\hline
		$\delta_{\mathrm{CP}}/^{\circ}$ & $-148_{-41}^{+26}$ & $[-180, 4] \cup [124, 180]$ \\
		\hline
		\hline
	\end{tabular}
	\caption{Best-fit values and their $3\sigma$ ranges for the neutrino oscillation parameters obtained from NuFIT 6.0~\cite{Esteban:2024eli}. Here LO and HO denote the lower and higher octants of $\theta_{23}$, respectively. The quoted range of $\delta_{\mathrm{CP}}$ is shown for reference and is not imposed as a constraint in the numerical scan.}
	\label{table:data_nufit}
\end{table}

\subsection{Simulation setup}
\label{simulation}

The experimental sensitivities of DUNE, T2HK, and JUNO are simulated using the GLoBES framework~\cite{Huber:2004ka,Huber:2007ji}. The configurations and exposures adopted for the three experiments are summarized below.\\

\noindent\textbf{\underline{DUNE:}} For the Deep Underground Neutrino Experiment (DUNE), we adopt the official configuration files associated with the Technical Design Report (TDR)~\cite{DUNE:2021cuw}. DUNE uses a high-intensity broadband neutrino beam generated at Fermilab, with a nominal beam power of 1.2~MW and a far detector located in South Dakota. The far detector consists of a 40~kt liquid argon time projection chamber (LArTPC), while the near-detector complex at Fermilab characterizes the un-oscillated beam.

For the DUNE analysis, we assume a total runtime of 10 years, divided equally between neutrino and antineutrino running modes, with 5 years in each mode. This corresponds to an annual exposure of $1.1 \times 10^{21}$ protons on target (POT). The adopted configuration follows the setup described in Ref.~\cite{DUNE:2020jqi}.\\

\noindent\textbf{\underline{T2HK:}} The Tokai to Hyper-Kamiokande (T2HK) experiment is a long-baseline accelerator experiment in Japan with a baseline of 295 km. To obtain a narrow-band energy spectrum, the simulated neutrino flux is evaluated at a $2.5^\circ$ off-axis angle. The neutrino beam generated at the J-PARC facility is assumed to operate with a power of 1.3 MW, corresponding to $2.7 \times 10^{21}$ POT per year. The far detector is taken to be a water Cherenkov detector with a fiducial mass of 374 kt. A total runtime of 10 years is assumed, distributed in a 1:3 ratio between neutrino and antineutrino modes. Detector systematics are implemented according to Refs.~\cite{Hyper-Kamiokande:2016srs, Hyper-Kamiokande:2018ofw}.\\

\noindent\textbf{\underline{JUNO:}} For the Jiangmen Underground Neutrino Observatory (JUNO), we use the experimental configuration described in Ref.~\cite{JUNO:2015zny}. The setup consists of a liquid scintillator detector with a 20 kt fiducial mass, positioned at a baseline of approximately 53 km from the Yangjiang and Taishan nuclear power plants. An energy resolution of $3\%/\sqrt{E \, (\text{MeV})}$ is applied to the detector model. For simplicity, all reactor cores are approximated as being located at an identical distance from the detector, and a total continuous runtime of 6 years is assumed.

The statistical component of the experimental sensitivity is evaluated using a $\chi^2$ function based on the Poisson log-likelihood ratio:
\begin{equation}
	\chi^2_{\rm stat} = 2 \sum_{i=1}^{n}
	\left[
	N^{\rm test}_i - N^{\rm true}_i
	- N^{\rm true}_i \ln\left(
	\frac{N^{\rm test}_i}{N^{\rm true}_i}
	\right)
	\right],
\end{equation}
where $N^{\rm true}_i$ and $N^{\rm test}_i$ represent the expected number of events in the $i^{\rm th}$ energy bin under the assumed true and test hypotheses, respectively, and $n$ denotes the total number of energy bins.

Systematic uncertainties are incorporated through the pull method~\cite{Fogli:2002pt,Huber:2002mx}, including signal and background normalization uncertainties together with shape uncertainties where applicable.

The oscillation parameters are taken from the latest global-fit results of \textsf{NuFIT}~6.0~\cite{Esteban:2024eli}, listed in Table~\ref{table:data_nufit}. During the statistical evaluation, we marginalize over the relevant oscillation parameters within the ranges permitted by the underlying theoretical model. The Dirac CP phase $\delta_{\rm CP}$ is varied over its entire physical range of $[-180^\circ,180^\circ]$ throughout the experimental simulations.

Furthermore, to evaluate the statistical agreement between our model predictions and the empirical \textsf{NuFIT}~6.0 results, we construct a supplementary $\chi^2$ function defined as
\begin{equation}
	\chi^2 = \sum_i
	\frac{\left(O^{\rm th}_i - O^{\rm exp}_i\right)^2}
	{\sigma_i^2},
	\label{chi-th}
\end{equation}
where $O^{\rm th}_i$ and $O^{\rm exp}_i$ denote the theoretical prediction and the experimental best-fit value of the $i^{\rm th}$ observable, respectively, and $\sigma_i$ denotes the corresponding experimental uncertainty. Each $O^{\rm th}_i$ depends on the underlying model parameters. The Dirac CP phase $\delta_{\rm CP}$ is not included in this $\chi^2$ function and is instead treated as a prediction of the model over its entire physical range. The numerical minimization of the $\chi^2$ function is performed using the Python package \textsf{iminuit} \cite{Dembinski_scikit-hep_iminuit}.

\section{Neutrino oscillation predictions}
\label{sec:results}

\subsection{Allowed parameter space and model constraints}

We identify the model parameter space consistent with the global neutrino oscillation data at the $3\sigma$ level summarized in Table~\ref{table:data_nufit}. The model contains five free parameters in addition to the complex modulus $\tau$. In the numerical scan, the complex modulus $\tau$ is restricted to the fundamental domain
\begin{equation}
	\mathcal{F} =
	\left\{
		\tau \in \mathbb{C} \;\bigg|\;
		\mathrm{Im}\,\tau > 0,\quad
		|\mathrm{Re}\,\tau| \leq \frac{1}{2},\quad
		|\tau| \geq 1
		\right\}.
\end{equation}
For numerical purposes, we further restrict the imaginary part to $\mathrm{Im}(\tau)\leq 5.0$, such that the scan is performed over
\begin{equation}
	-\frac{1}{2}\leq\mathrm{Re}(\tau)\leq\frac{1}{2},
	\qquad
	0.86\leq\mathrm{Im}(\tau)\leq5.0,
	\qquad
	|\tau|\geq1.
\end{equation}
The dimensionless coupling ratios are varied over
\begin{equation}
	\tilde\beta_{D},\tilde\beta_{RS},\tilde\gamma_{D},\tilde\gamma_{RS},\tilde\gamma_{S}
	\in \left[10^{-8},10^4\right].
\end{equation}

For each parameter point, the overall coefficients $\alpha_D$, $\alpha_{RS}$, and $\alpha_S$ are chosen such that the corresponding dimensionless coupling combinations, including the modular-form contributions, remain of order unity or smaller.\footnote{For example, from Eq.~\eqref{eq:dirac_neutrino_mass_matrix}, the entries of the Dirac-neutrino Yukawa matrix contain the combinations $\alpha_DY_{3,i}^{(-2)}$, $\beta_DY_{3,i}^{(-2)}$, and $\gamma_DY_{3,i}^{(-2)}$. We impose the conservative order-one perturbative criterion $|\alpha Y|,|\beta Y|,|\gamma Y|\lesssim1$ on the analogous dimensionless combinations entering the different mass sectors \cite{Bernal:2021kaj,Ipek:2018sai}. Since the remaining couplings are related through $\beta=\alpha\widetilde{\beta}$ and $\gamma=\alpha\widetilde{\gamma}$, the imposed condition ensures that these dependent couplings remain within the adopted perturbative range.}

Under these conditions, the parameter space compatible with the neutrino oscillation constraints yields the following range for the sum of the light-neutrino masses:
\begin{equation}
	\sum_i m_i \in \left[0.059,0.684\right]~\mathrm{eV}.
\end{equation}
Cosmological constraints on the absolute neutrino-mass scale depend on the adopted cosmological model and the combination of datasets employed \cite{Planck:2018vyg,DiValentino:2019dzu}. In the present analysis, we adopt the Planck 2018 constraint obtained in combination with BAO measurements,
\begin{equation}
	\sum_i m_i < 0.12~\mathrm{eV},
\end{equation}
at $95\%$ C.L. \cite{Planck:2018vyg}. Accordingly, this cosmological bound is imposed when selecting parameter points for the subsequent leptogenesis analysis.

\begin{figure}[t]
	\centering
	\includegraphics[width=0.75\linewidth]{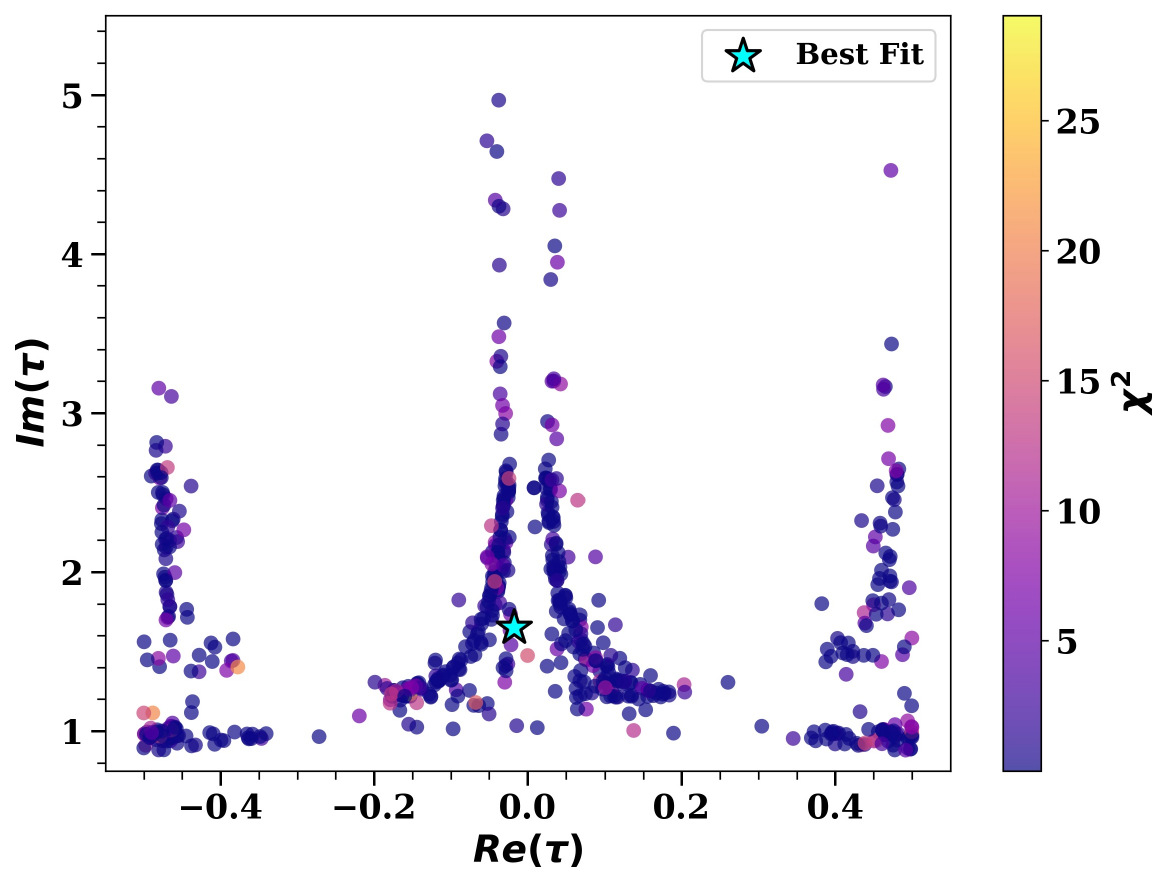}
	\caption{Correlation between $Re(\tau)$ and $Im(\tau)$. The blue star represents the best-fit value.}
	\label{fig:tau}
\end{figure}

Within the parameter ranges considered, our model predicts only Normal Ordering (NO) for phenomenologically viable neutrino masses and mixing. The corresponding allowed region in the complex $\tau$ plane is shown in Fig.~\ref{fig:tau}. From the numerical scan, the best-fit point is obtained at $\mathrm{Re}(\tau)=-0.018$ and $\mathrm{Im}(\tau)=1.65$. This best-fit point is indicated by a blue star in the figure. The corresponding best-fit values and scan ranges for the complex modulus are summarized in Table~\ref{tab:best-fit-param}.

\begin{table}[t]
	\centering
	\renewcommand{\arraystretch}{1.1}
	\setlength{\tabcolsep}{4pt}
	\small
	\begin{tabular}{|c|c|c|}
		\hline
		\multirow{2}{*}{\textbf{Parameters}}
		& \multicolumn{2}{c|}{\textbf{Normal Ordering (NO)}} \\ \cline{2-3}
		& Best fit & Scan range \\ \hline
		Re($\tau$) &$-0.018$ & [-0.5, 0.5] \\ \hline
		Im ($\tau$) &$1.65$& [0.86, 5] \\ \hline
	\end{tabular}
	\caption{Best-fit values and scan ranges for the complex modulus parameter under NO.}
	\label{tab:best-fit-param}
\end{table}

\subsection{Projected sensitivities of DUNE, T2HK, and JUNO}

\begin{figure}[t]
	\centering
	\includegraphics[width=0.47\linewidth]{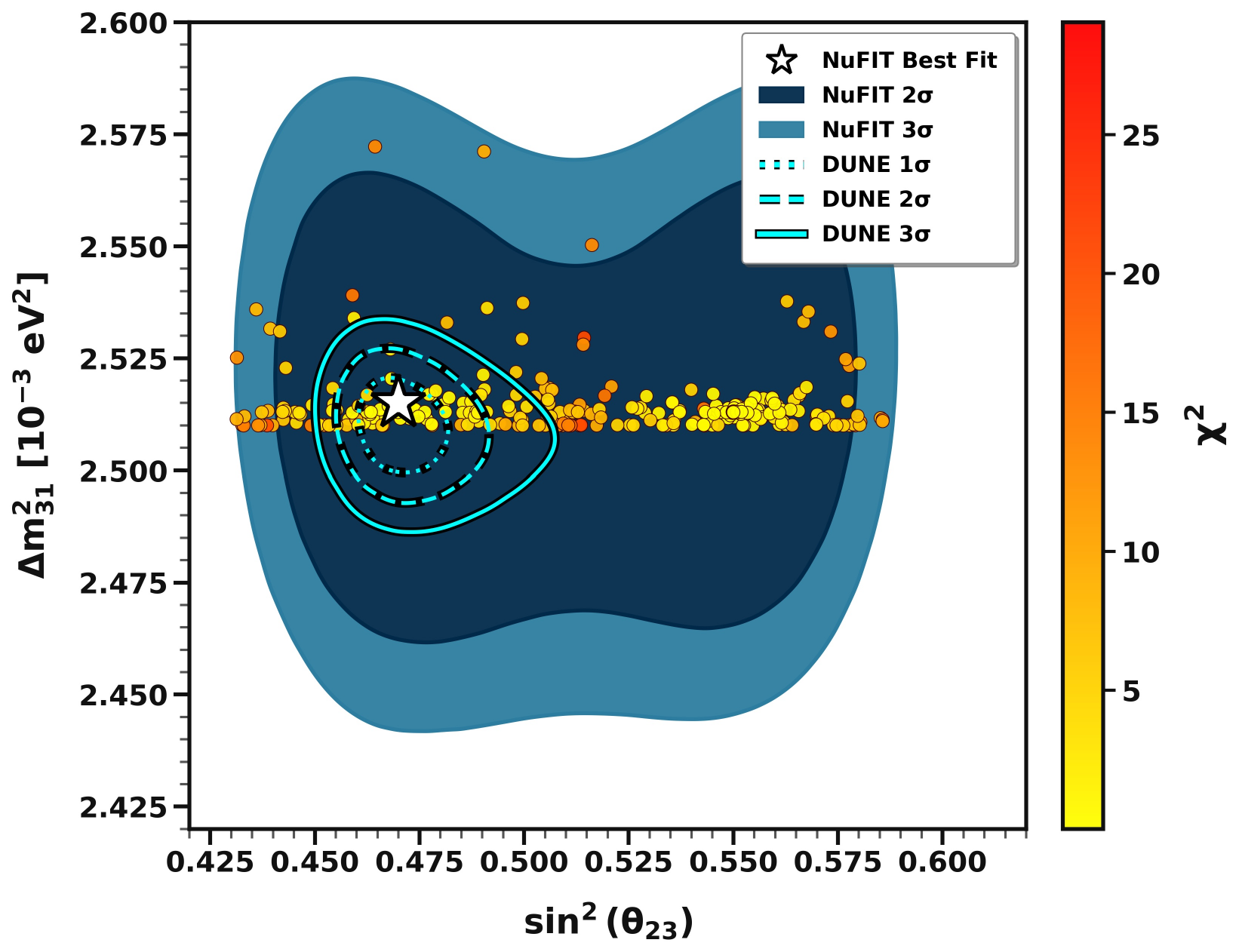}
	\includegraphics[width=0.47\linewidth]{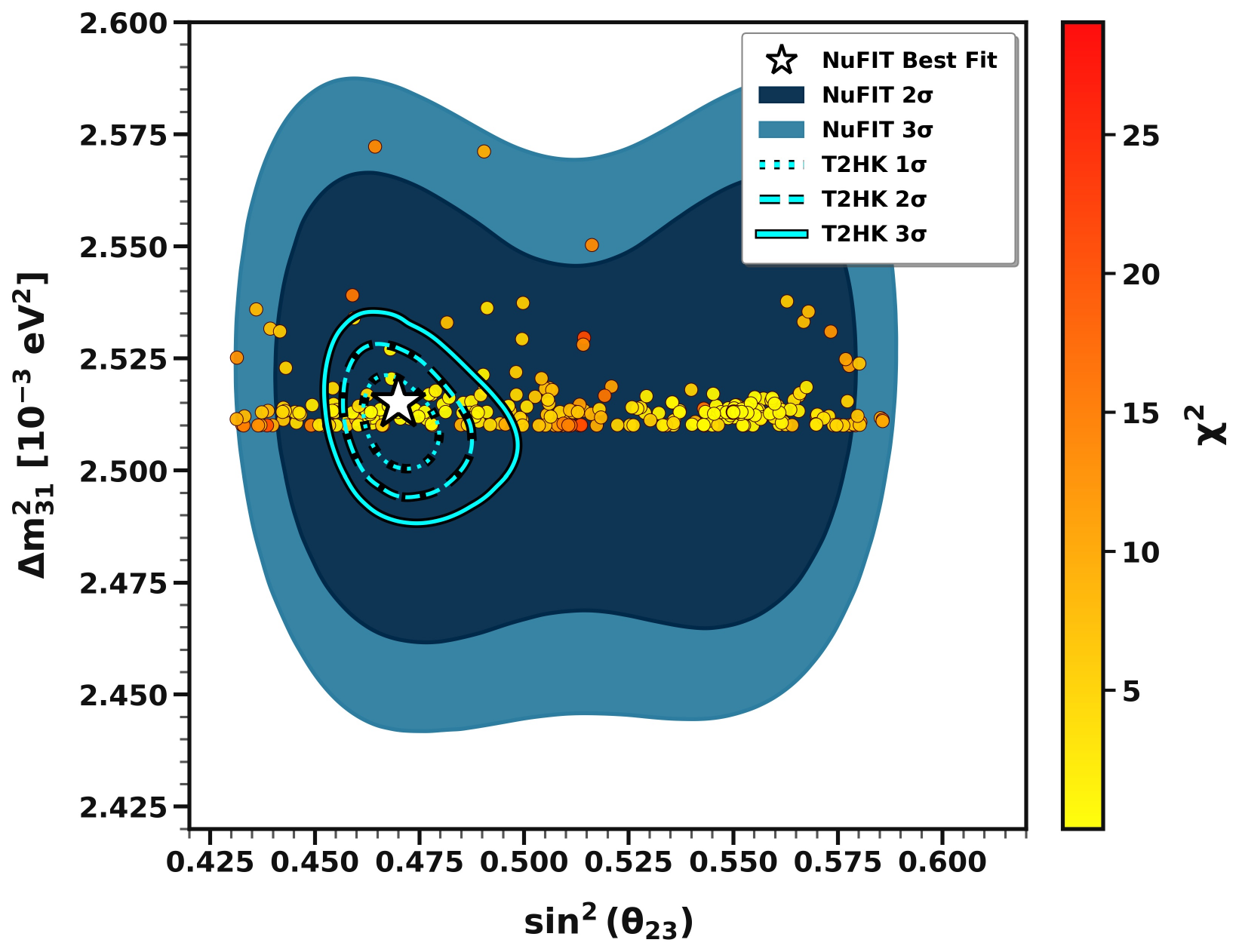}
	\includegraphics[width=0.47\linewidth]{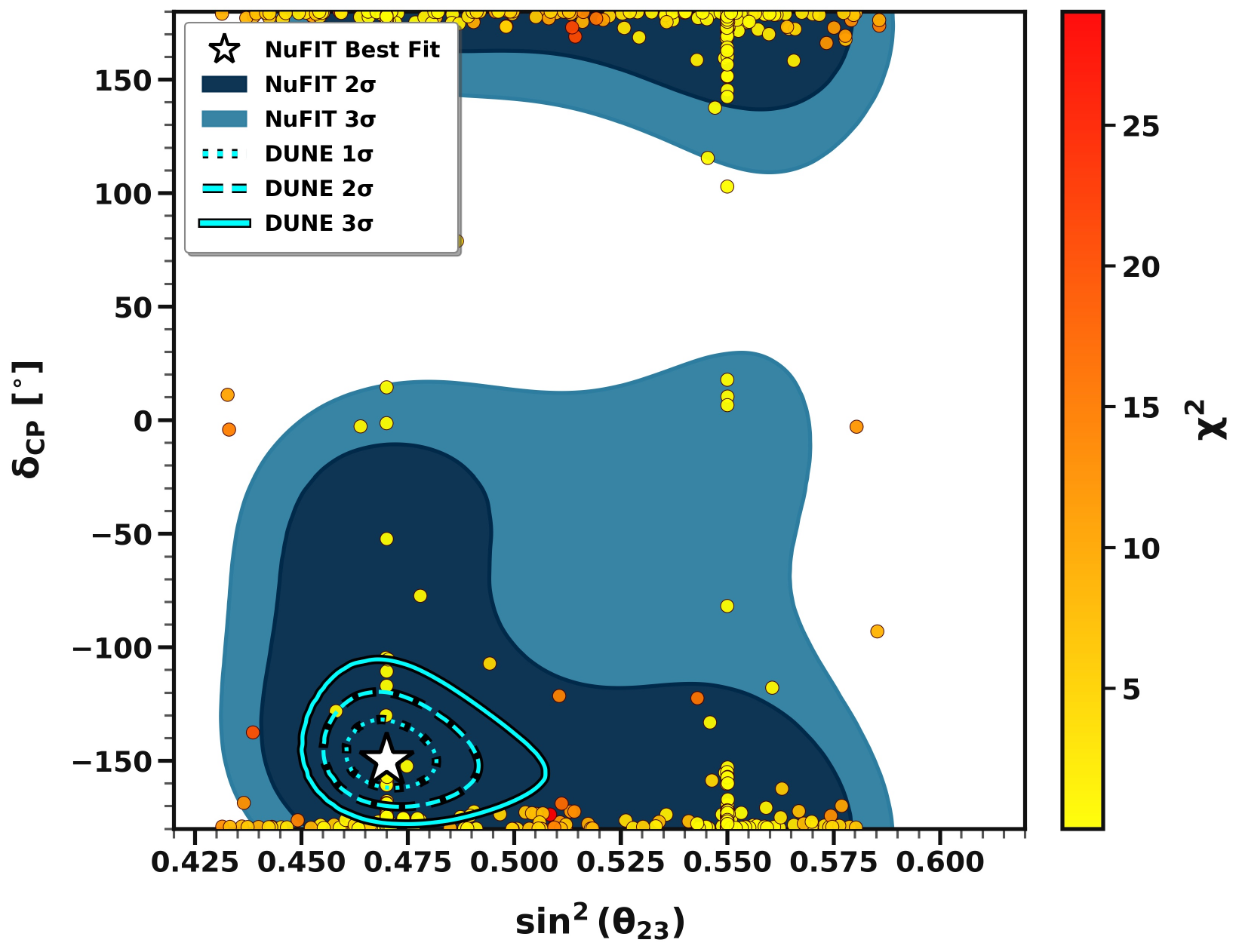}
	\includegraphics[width=0.47\linewidth]{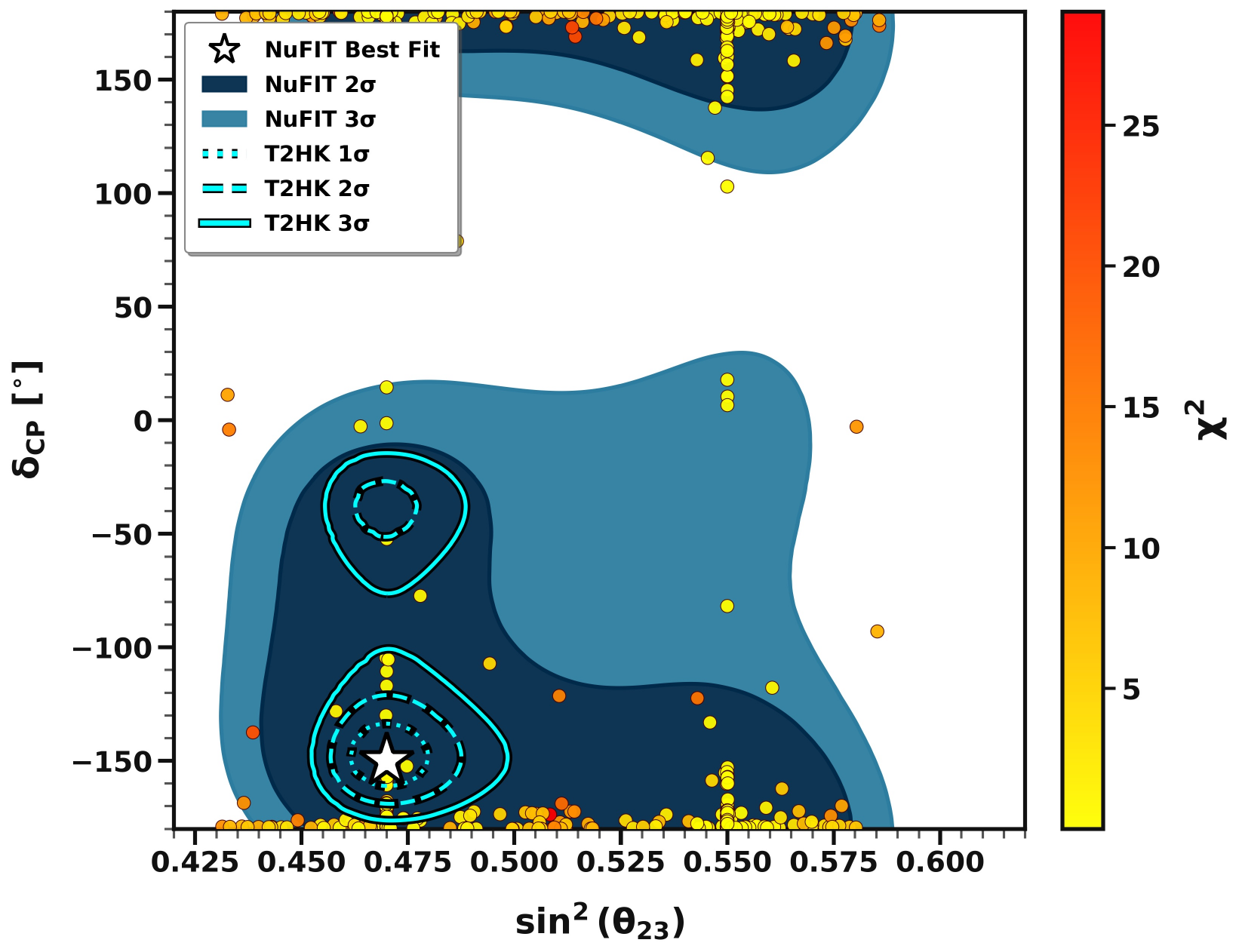}
	\includegraphics[width=0.47\linewidth]{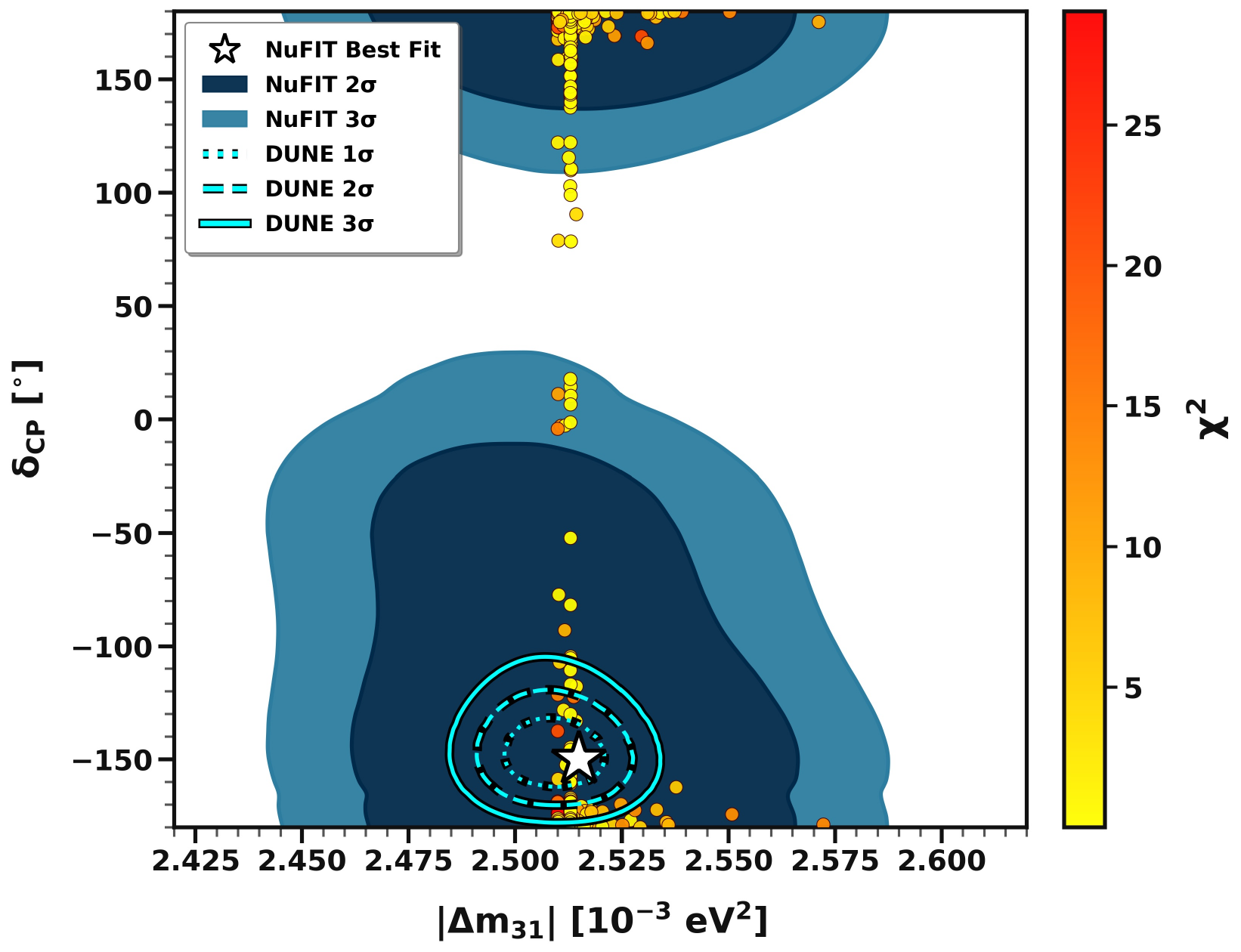}
	\includegraphics[width=0.47\linewidth]{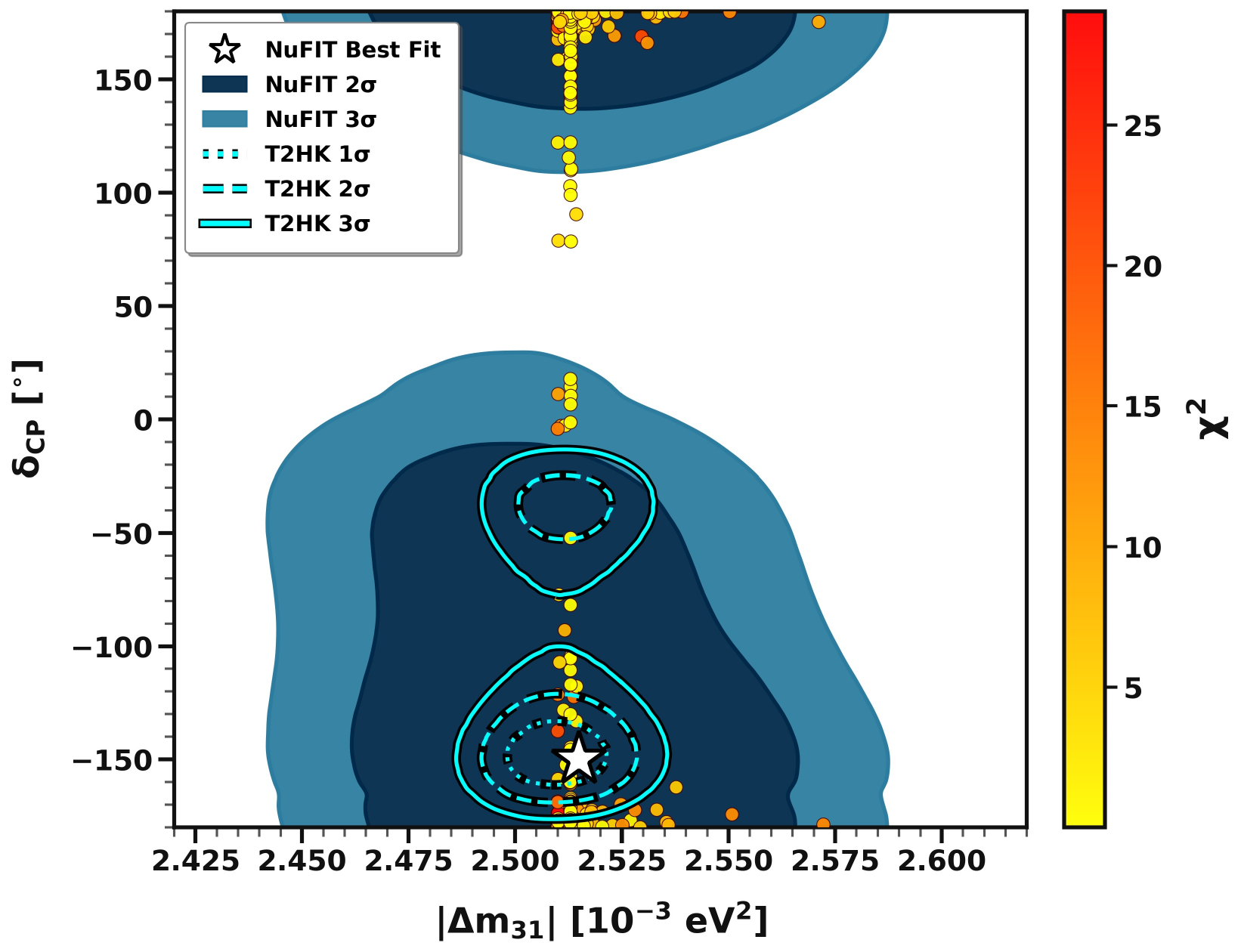}
	\caption{Correlations among the oscillation parameters $\Delta m^2_{31}$, $\delta_{\rm CP}$, and $\theta_{23}$ for normal ordering. The projected sensitivity contours correspond to DUNE (left panels) and T2HK (right panels), obtained assuming the lower-octant best-fit value of $\theta_{23}$ as the true hypothesis, while the green contours indicate the current \textsf{NuFIT} constraints. The scattered points represent the parameter space allowed by the model.}
	\label{fig:Dune-t2hk1}
\end{figure}

\begin{figure}[t]
	\centering
	\includegraphics[width=0.47\linewidth]{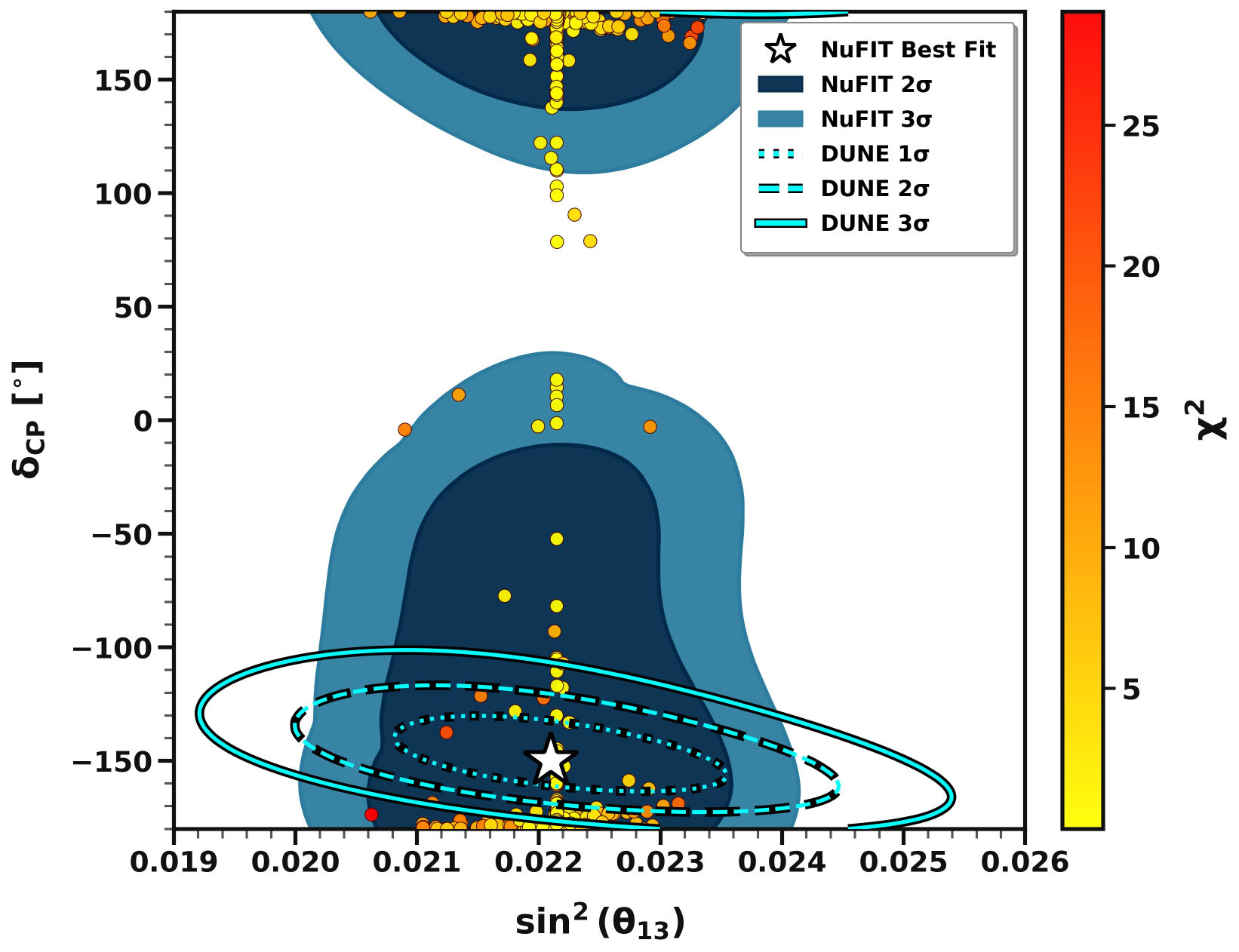}
	\includegraphics[width=0.47\linewidth]{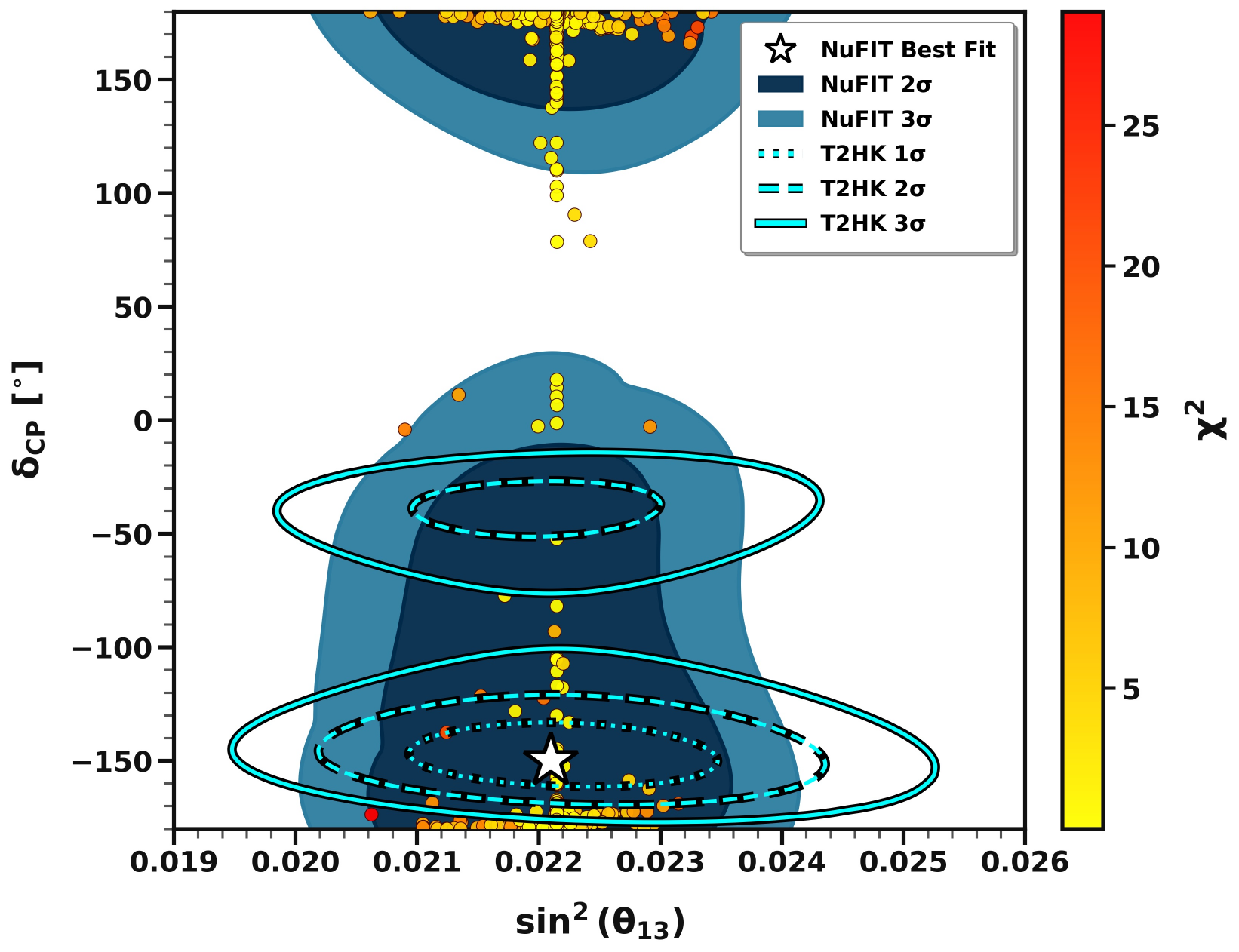}
	\includegraphics[width=0.47\linewidth]{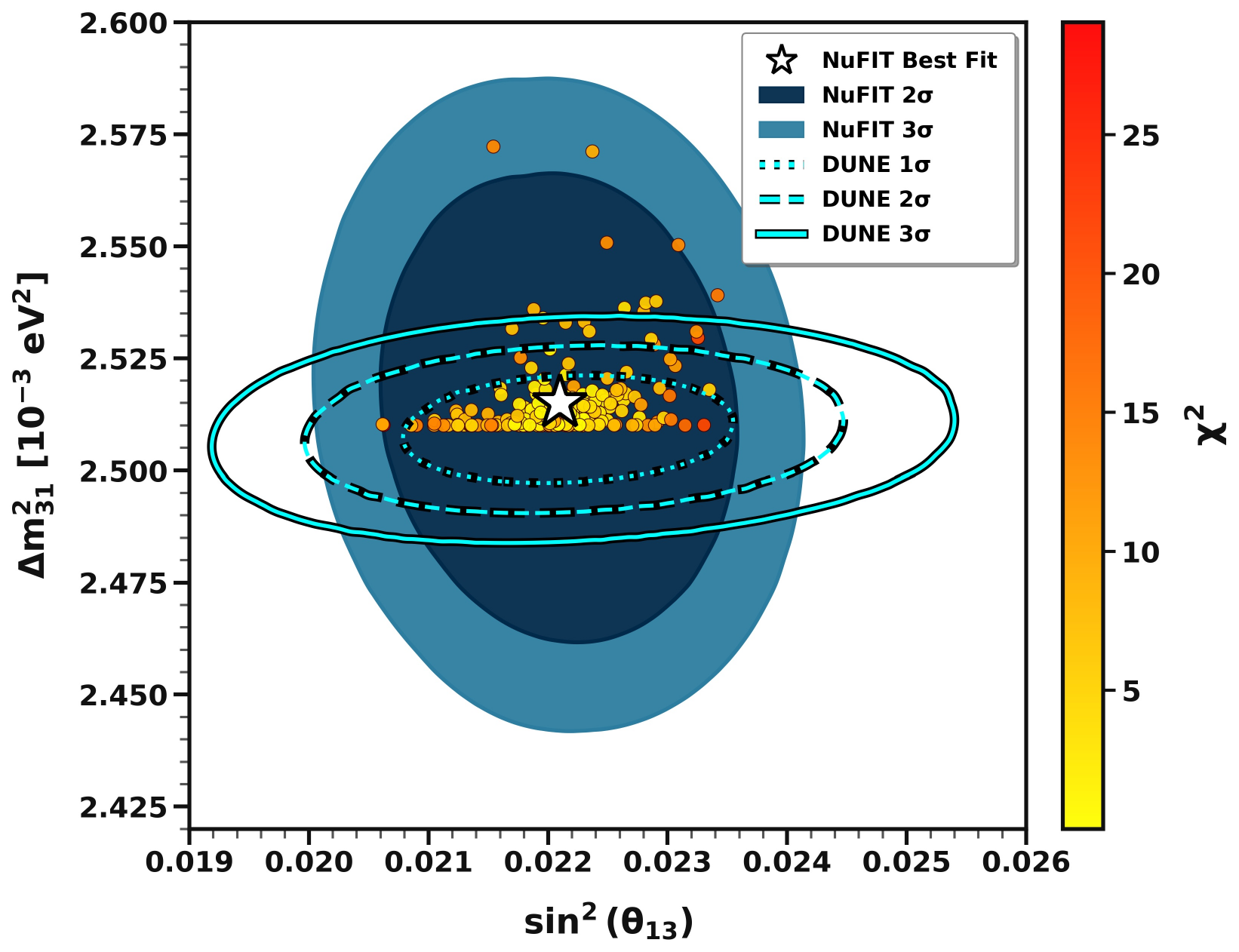}
	\includegraphics[width=0.47\linewidth]{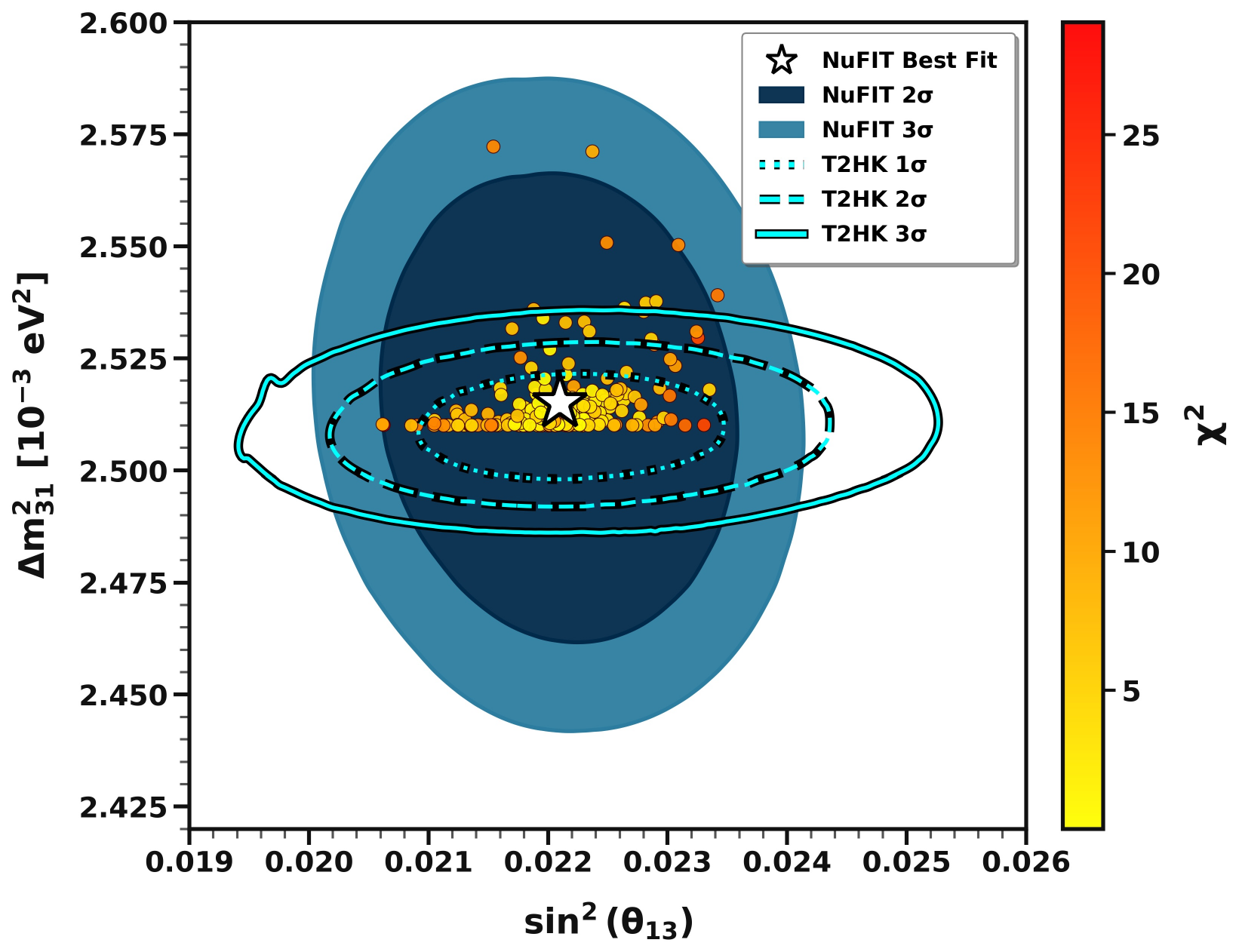}
	\includegraphics[width=0.47\linewidth]{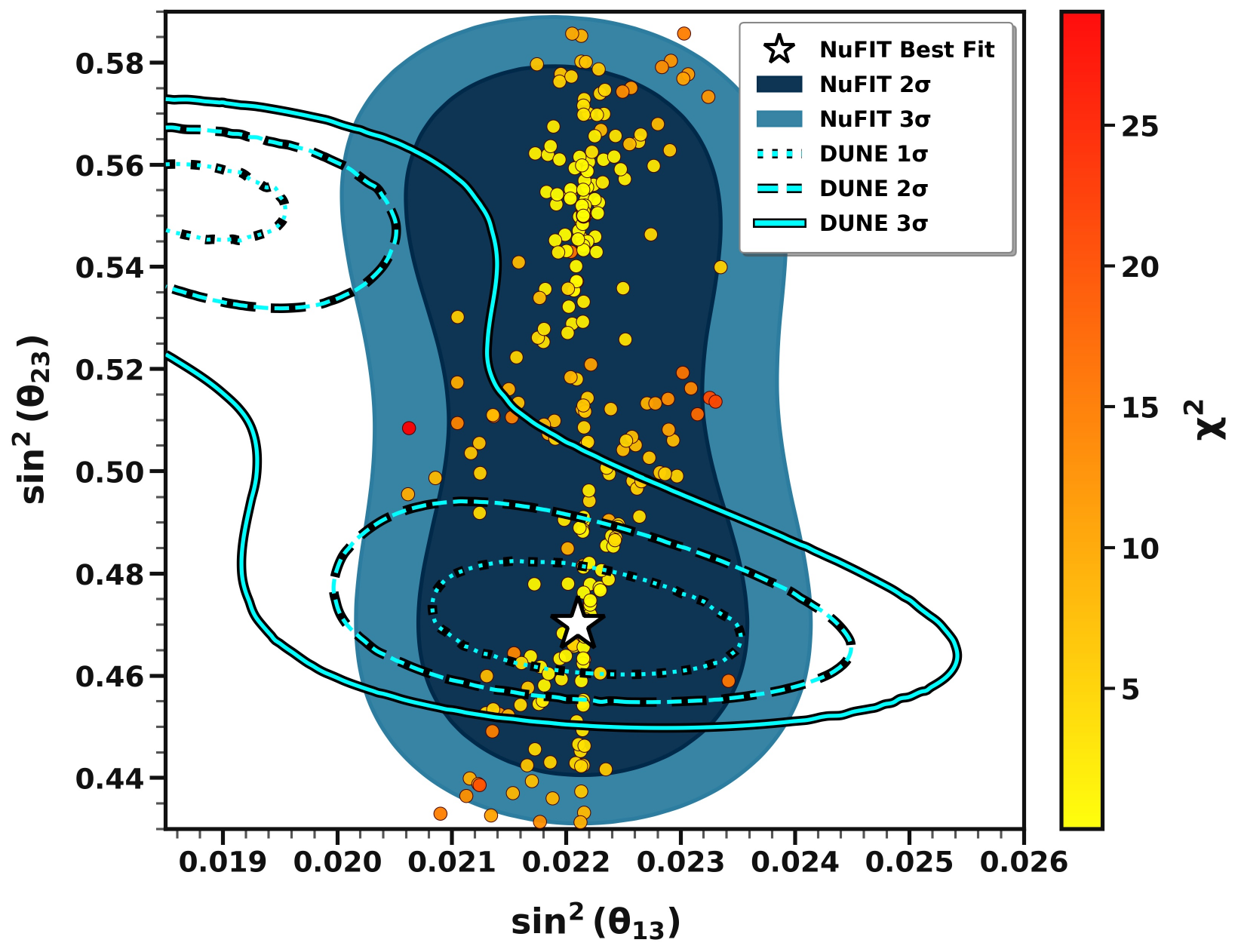}
	\includegraphics[width=0.47\linewidth]{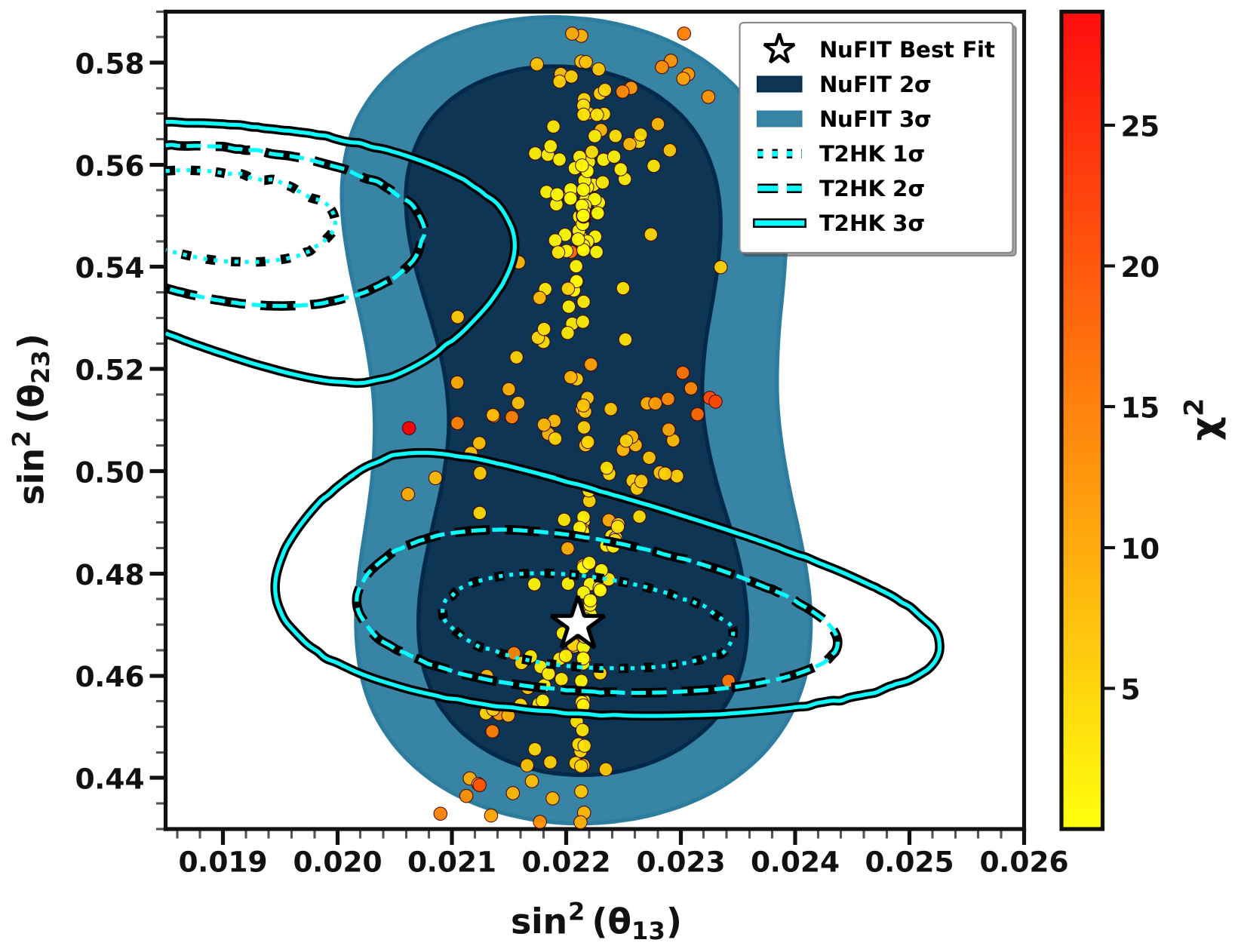}
	\caption{Correlations among the oscillation parameters $\Delta m^2_{31}$, $\delta_{\rm CP}$, $\theta_{13}$, and $\theta_{23}$ for normal ordering. The projected sensitivity contours correspond to DUNE (left panels) and T2HK (right panels), obtained assuming the lower-octant best-fit value of $\theta_{23}$ as the true hypothesis, while the green contours indicate the current \textsf{NuFIT} constraints. The scattered points represent the parameter space allowed by the model.}
	\label{fig:Dune-t2hk2}
\end{figure}

\begin{figure}[t]
	\centering
	\includegraphics[width=0.47\linewidth]{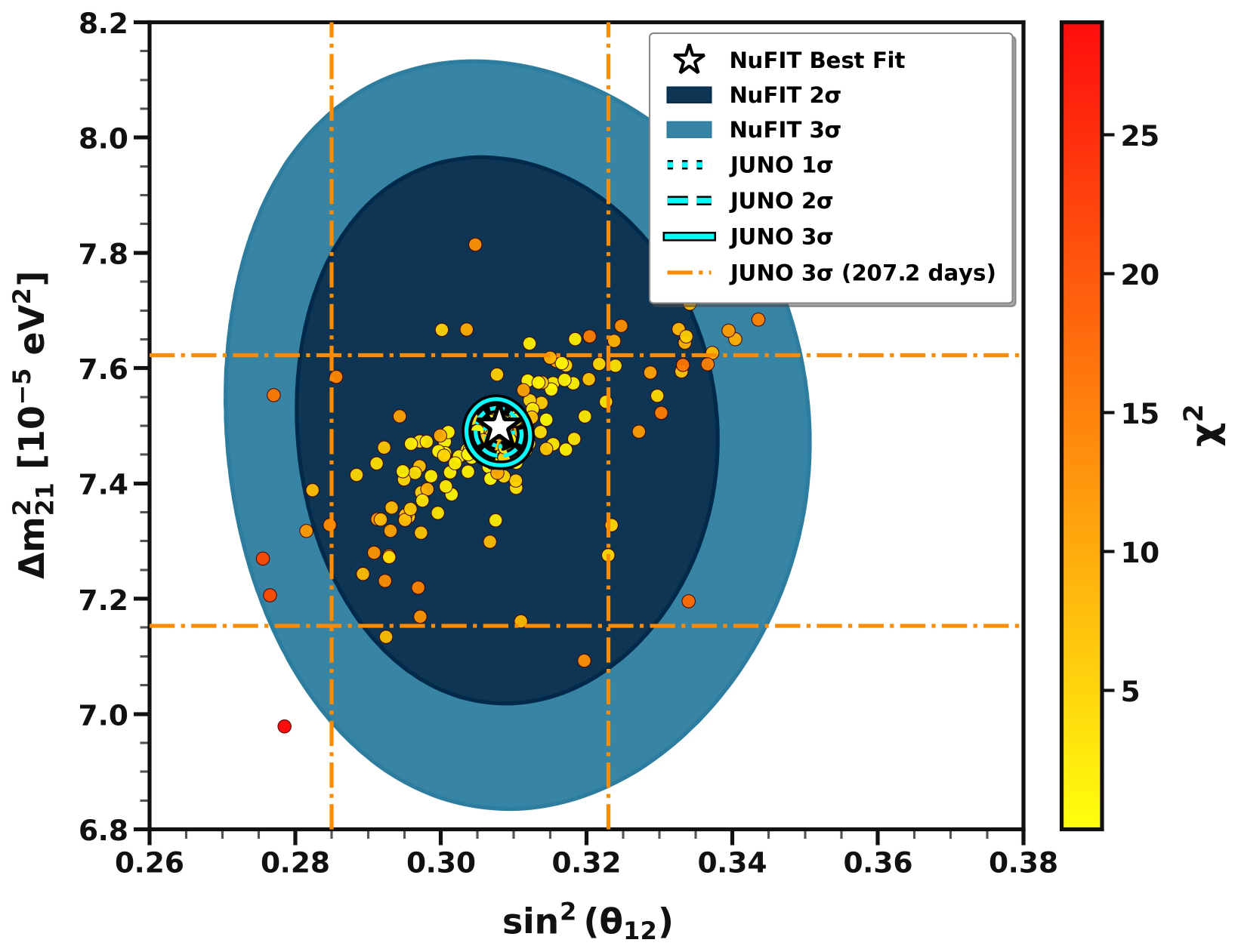}
	\includegraphics[width=0.47\linewidth]{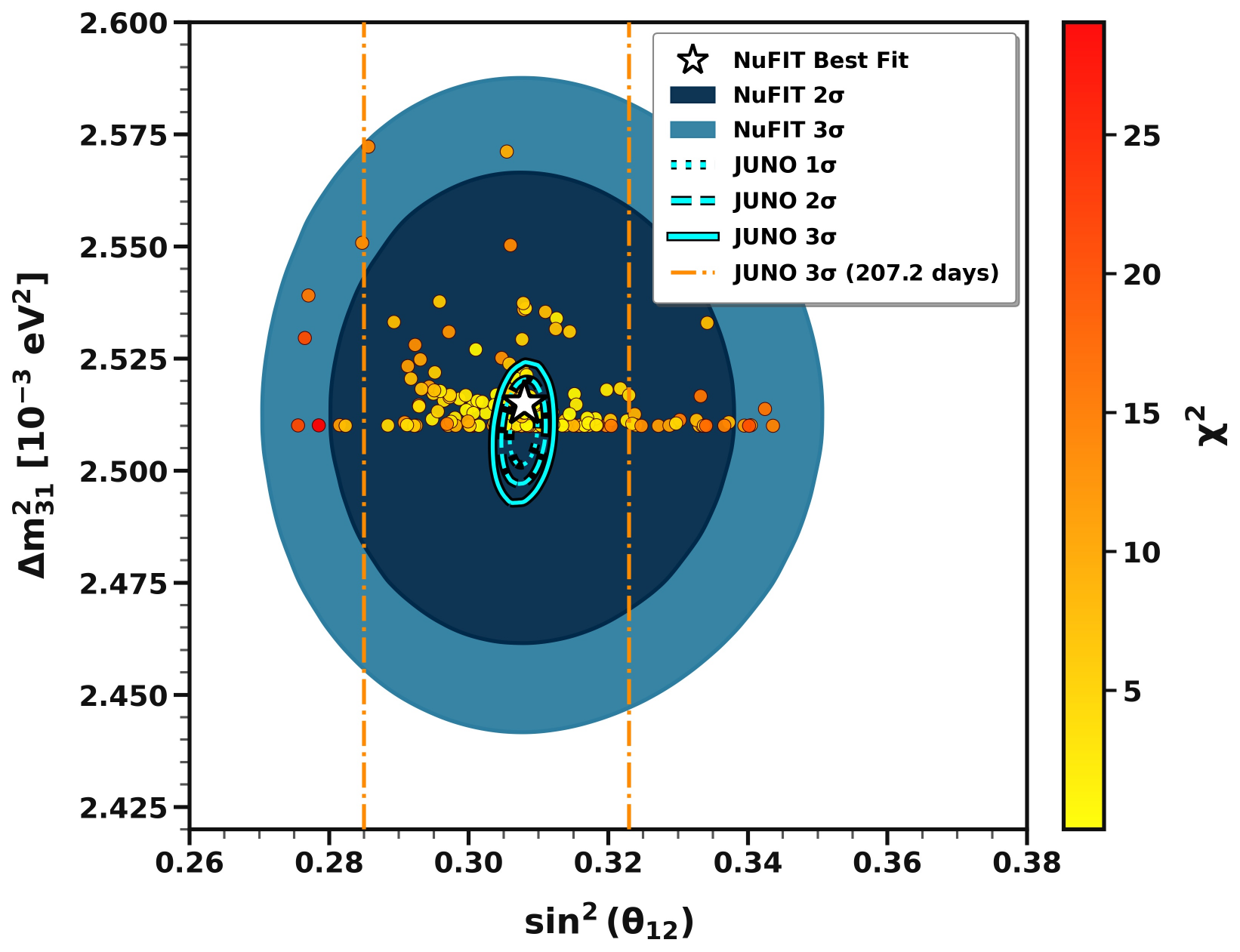}
	\caption{Correlations among the oscillation parameters $\Delta m^2_{31}$, $\Delta m^2_{21}$, and $\theta_{12}$ for normal ordering. The projected sensitivity contours correspond to JUNO, while the green contours indicate the current \textsf{NuFIT} constraints. The vertical dot-dashed lines in both panels represent the current $3\sigma$ JUNO constraint on $\sin^2\theta_{12}$ obtained from 207.2 days of data taking, while the horizontal dot-dashed lines in the left panel indicate the corresponding current $3\sigma$ constraint on $\Delta m^2_{21}$. The current JUNO constraint on $\Delta m^2_{31}$ is not shown in the right panel. The scattered points represent the parameter space allowed by the model.}
	\label{fig:Juno}
\end{figure}

By scanning the allowed model parameter space, we obtain the corresponding predictions for the neutrino oscillation observables. Figures~\ref{fig:Dune-t2hk1} and \ref{fig:Dune-t2hk2} display the correlations among $\sin^2 \theta_{13}$, $\sin^2 \theta_{23}$, $\Delta m^2_{31}$, and $\delta_{\rm CP}$ together with the projected sensitivities of DUNE (left panels) and T2HK (right panels). Similarly, Fig.~\ref{fig:Juno} shows the JUNO sensitivities involving $\sin^2 \theta_{12}$, $\Delta m^2_{21}$, and $\Delta m^2_{31}$.

The allowed regions for DUNE and T2HK are obtained by marginalizing over the relevant oscillation parameters $\theta_{23}$, $\theta_{13}$, $\Delta m^2_{31}$, and $\delta_{\rm CP}$, except for the parameters explicitly shown in a given plane. The Dirac CP phase $\delta_{\rm CP}$ is varied over its entire physical range of $[-180^\circ,180^\circ]$ throughout the simulations, either as a plotted parameter or as a marginalized parameter when it is not displayed. In these plots, the scattered points represent the parameter space allowed by our model, while the color bars indicate the corresponding $\chi^2$ values. The solid and dashed contours depict the projected sensitivities of the respective experiments at the $1\sigma$, $2\sigma$, and $3\sigma$ confidence levels, while the green shaded bands indicate the $2\sigma$ and $3\sigma$ regions allowed by \textsf{NuFIT}. Overall, our model accommodates $\delta_{\rm CP}$ across its entire physical range, while the predicted values of $\sin^2 \theta_{13}$, $\sin^2 \theta_{12}$, $\Delta m^2_{21}$, and $\Delta m^2_{31}$ remain within the regions allowed by \textsf{NuFIT}.

Figure~\ref{fig:Dune-t2hk1} shows the model predictions in the planes involving $\sin^2\theta_{23}$, $\delta_{\rm CP}$, and the atmospheric mass-squared difference $\Delta m^2_{31}$, together with the projected sensitivities of DUNE and T2HK. In generating these precision contours, the lower-octant NuFIT best-fit value of $\theta_{23}$ is taken as the true input, while the test parameter space is scanned over the $3\sigma$ allowed range (includes both the lower and higher octants). This allows the possible octant degeneracy to be examined within the projected experimental sensitivities. The white stars indicate the \textsf{NuFIT} best-fit values used as reference points in the simulations. The individual correlations shown in Fig.~\ref{fig:Dune-t2hk1} are discussed in detail below.\\

\noindent \textbf{\underline{Top Panels ($\bm{\sin^2\theta_{23}}$ vs. $\bm{\Delta m^2_{31}}$):}}\\
Compared with the broad \textsf{NuFIT} constraints, which currently allow both octants of $\theta_{23}$, the projected DUNE and T2HK contours substantially reduce the allowed parameter space in the $\sin^2\theta_{23}$-$\Delta m^2_{31}$ plane. Under the lower-octant true hypothesis adopted in the precision simulations, the $1\sigma$, $2\sigma$, and $3\sigma$ sensitivity regions of both experiments remain exclusively localized around the lower-octant best-fit point ($\sin^2\theta_{23}\approx0.47$), with no surviving upper-octant degeneracy at the $3\sigma$ level. Within this simulated scenario, DUNE and T2HK can therefore resolve the $\theta_{23}$ octant degeneracy while providing a precise determination of $\Delta m^2_{31}$. A dense horizontal band of highly favored, low-$\chi^2$ model predictions shows substantial overlap with these stringent lower-octant experimental projections. Consequently, if nature favors the lower octant, these next-generation experiments will not only tightly constrain $\Delta m^2_{31}$ but can also exclude at the $3\sigma$ level the higher-octant region currently populated by a subset of our model predictions.\\

\noindent \textbf{\underline{Middle Panels ($\bm{\sin^2\theta_{23}}$ vs. $\bm{\delta_{\rm CP}}$):}}\\
The projected sensitivities of DUNE (left panel) and T2HK (right panel) substantially narrow the broad parameter space currently allowed by \textsf{NuFIT} in the $\delta_{\rm CP}$ versus $\sin^2\theta_{23}$ plane. Under the lower-octant true hypothesis adopted in the precision simulations, the projected sensitivities of both experiments strongly localize the allowed parameter space to the lower octant near $\sin^2\theta_{23}\approx0.47$, with a preferred Dirac CP phase around $-150^\circ$, and resolve the octant degeneracy up to the $3\sigma$ confidence level. T2HK, however, exhibits an additional degenerate solution in $\delta_{\rm CP}$ near $-50^\circ$ at the $2\sigma$ and $3\sigma$ levels. The theoretical model predictions show strong compatibility with these projected constraints, with the highly favored, low-$\chi^2$ points clustering densely within the lower-octant sensitivity regions. A pronounced accumulation of model predictions is also observed near the CP-conserving values $\delta_{\rm CP}=\pm180^\circ$.\\

\noindent \textbf{\underline{Lower Panels ($\bm{\Delta m^2_{31}}$ vs. $\bm{\delta_{\rm CP}}$):}}\\
The lower panels explore the $\Delta m^2_{31}$ versus $\delta_{\rm CP}$ parameter space, demonstrating how both DUNE and T2HK will drastically refine the broad bounds currently allowed by \textsf{NuFIT}. While our model accommodates the entire physical range of $\delta_{\rm CP}$, it strictly restricts $\Delta m^2_{31}$ to a narrow interval of $[2.5, 2.58] \times 10^{-3} \text{ eV}^2$, fully consistent with \textsf{NuFIT} constraints. Comparing the two long-baseline options, DUNE provides superior constraints on $\delta_{\rm CP}$ by avoiding the aforementioned parameter degeneracy that afflicts T2HK. Overall, the projected experimental contours tightly bound these theoretical predictions, with the vast majority of the highly favored, low $\chi^2$ model points falling well within the anticipated sensitivities of both the experiments.

Turning to correlations involving $\theta_{13}$, Fig.~\ref{fig:Dune-t2hk2} shows the projected sensitivities in the planes involving $\sin^2\theta_{13}$, $\Delta m^2_{31}$, $\delta_{\rm CP}$, and $\sin^2\theta_{23}$. A notable feature across these panels is the comparatively weaker sensitivity of the long-baseline experiments DUNE and T2HK to $\sin^2\theta_{13}$ than the present global \textsf{NuFIT} constraint, as reflected by the broad, elongated projected contours along the $\sin^2\theta_{13}$ direction. This reduced precision arises from persistent parameter degeneracies in long-baseline measurements, while reactor experiments such as Daya Bay, RENO, and Double Chooz continue to provide considerably stronger constraints on $\theta_{13}$. Despite the broad contours of the long-baseline projections, our underlying theoretical model restricts $\sin^2\theta_{13}$ to the narrow, \textsf{NuFIT}-favored interval of approximately $[0.020,0.024]$. In the top and middle panels, the highly favored, low-$\chi^2$ model predictions show strong overlap with the regions where the \textsf{NuFIT} constraints intersect the projected DUNE and T2HK sensitivities for $\delta_{\rm CP}$ and $\Delta m^2_{31}$.

The lower panels, showing the $\sin^2\theta_{23}$ versus $\sin^2\theta_{13}$ plane, reveal a more distinctive separation between the two octant regions. Under the lower-octant true hypothesis adopted in the precision simulations, the model predictions near $\sin^2\theta_{23}\approx0.47$ remain consistent with both the present \textsf{NuFIT} region and the primary DUNE and T2HK sensitivity contours. Both experiments also exhibit degenerate sensitivity islands in the higher octant; however, these regions are shifted toward smaller values of $\sin^2\theta_{13}$. In contrast, the higher-octant model solutions near $\sin^2\theta_{23}\approx0.55$ remain confined to the higher, \textsf{NuFIT}-favored $\sin^2\theta_{13}$ region and therefore do not overlap with these projected upper-octant islands, showing tension with the projected contours from the $1\sigma$ through the $3\sigma$ levels. Consequently, if the lower octant is realized in nature, the projected DUNE and T2HK measurements can discriminate against the higher-octant solutions presently allowed by the model.

Since DUNE and T2HK have comparatively limited sensitivity to the solar oscillation parameters, we do not consider their contours in parameter planes involving $\Delta m^2_{21}$. Instead, we examine the corresponding sensitivities of JUNO in Fig.~\ref{fig:Juno}, which is designed to provide high-precision measurements of both the solar mass-squared difference $\Delta m^2_{21}$ and the mixing angle $\theta_{12}$. The model constrains $\Delta m^2_{21}$ to approximately $[7.0,7.8]\times10^{-5}~\mathrm{eV}^2$ and $\sin^2\theta_{12}$ to $[0.27,0.35]$, within the presently allowed \textsf{NuFIT} ranges. Owing to JUNO's strong sensitivity to these solar oscillation parameters, its future full-exposure measurements will considerably narrow the current allowed regions. Since the model predictions span a substantially broader range in both $\sin^2\theta_{12}$ and $\Delta m^2_{21}$, only a localized subset of the presently viable model points overlaps with the projected $1\sigma$ to $3\sigma$ regions. JUNO can therefore provide a stringent test of the model predictions in the solar-neutrino sector and significantly restrict the currently allowed parameter space.

Recent JUNO results presented at Neutrino 2026 further demonstrate its strong sensitivity in the solar-neutrino sector~\cite{Yifang:2026}. In the left panel of Fig.~\ref{fig:Juno}, the vertical and horizontal dot-dashed lines indicate the current $3\sigma$ constraints on $\sin^2\theta_{12}$ and $\Delta m^2_{21}$, respectively, obtained from 207.2 days of data taking~\cite{Yifang:2026}. Despite the limited exposure, these constraints already exclude a substantial portion of the parameter region presently allowed by \textsf{NuFIT}. Beyond the solar sector, JUNO also exhibits strong projected sensitivity to the atmospheric mass-squared difference $\Delta m^2_{31}$, as shown in the right panel of Fig.~\ref{fig:Juno}, with the projected precision in our simulation exceeding that of the long-baseline experiments considered above. The corresponding constraint from the current 207.2-day JUNO dataset is not yet competitive with the established \textsf{NuFIT} range and has therefore been omitted from the figure.

Taken together, within the parameter ranges considered, our model predicts only Normal Ordering for phenomenologically viable neutrino masses and mixing, while $\delta_{\rm CP}$ spans its entire physical range. The projected sensitivities of DUNE, T2HK, and JUNO, together with emerging JUNO data, can further test and constrain complementary regions of the currently allowed model parameter space.
%%%%%%%%%%%%%%%%%%%%%%%%%%%%%%%%%%%%%%%%%%%%%%%%%%%%%%%%%
\section{Leptogenesis}\label{sec:leptogenesis}
\begin{figure}[t]
	\centering
	\resizebox{0.98\textwidth}{!}{%
		\begin{tikzpicture}[line width=1pt, scale=1.0]
			%================================================
			% Tree-level diagram
			%================================================
			\begin{scope}[shift={(0,0)}]
				\draw[color=black,solid] (-2.0,0.0)--(0.0,0.0);
				\draw[color=blue,dashed] (0.0,0.0)--(2.0,1.0);
				\draw[color=black,solid] (0.0,0.0)--(2.0,-1.0);
				
				\node at (-2.35,0.0) {$N_i$};
				\node[above] at (2.20,0.75) {$H$};
				\node[below] at (2.25,-0.70) {$\ell_\alpha$};
				\node at (0,-1.55) {(a)};
			\end{scope}
			
			%================================================
			% Self-energy one-loop diagram
			%================================================
			\begin{scope}[shift={(5.2,0)}]
				\draw[color=black,solid] (-2.0,0.0)--(-1.3,0.0);
				\draw[color=black,solid] (-0.1,0.0)--(0.6,0.0);
				\draw[color=blue,dashed] (0.6,0.0)--(2.1,1.0);
				\draw[color=black,solid] (0.6,0.0)--(2.1,-1.0);
				
				\draw[color=blue,dashed] (-1.3,0) arc (180:0:0.6cm);
				\draw[color=black,solid] (-1.3,0) arc (-180:0:0.6cm);
				
				\node at (-2.35,0.0) {$N_i$};
				\node[above] at (0.25,-0.10) {$N_j$};
				\node[above] at (2.28,0.75) {$H$};
				\node[below] at (2.28,-0.75) {$\ell_\alpha$};
				\node[above] at (-0.7,0.55) {$H$};
				\node[below] at (-0.7,-0.60) {$\ell,\bar{\ell}$};
				\node at (0,-1.55) {(b)};
			\end{scope}
			
			%================================================
			% Vertex one-loop diagram
			%================================================
			\begin{scope}[shift={(10.4,0)}]
				\draw[color=black,solid] (-2.0,0.0)--(-1.0,0.0);
				\draw[color=black,solid] (-1.0,0.0)--(0.6,1.0);
				\draw[color=blue,dashed] (-1.0,0.0)--(0.6,-1.0);
				\draw[color=blue,dashed] (0.6,1.0)--(2.0,1.0);
				\draw[color=black,solid] (0.6,-1.0)--(0.6,1.0);
				\draw[color=black,solid] (0.6,-1.0)--(2.0,-1.0);
				
				\node at (-2.35,0.0) {$N_i$};
				\node at (1.0,0.0) {$N_j$};
				\node[above] at (2.15,0.65) {$H$};
				\node[below] at (2.15,-0.65) {$\ell$};
				\node[above] at (-0.25,-1.15) {$H$};
				\node[below] at (-0.25,1.15) {$\ell$};
				\node at (0,-1.55) {(c)};
			\end{scope}
			
		\end{tikzpicture}
	}
	\caption{Tree-level, self-energy one-loop, and vertex one-loop diagrams for heavy neutrino decay. The asymmetry $\epsilon_i$ arises from the interference of the one-loop diagrams with the tree-level amplitude, with $j\neq i$ running inside the loop.}
	\label{fig:leptogenesis_diagram}
	\end{figure}
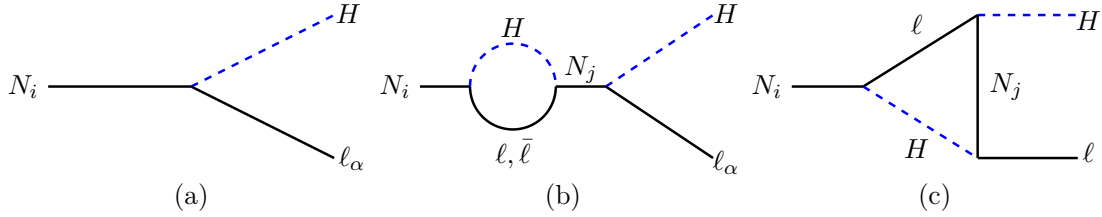
The present double seesaw framework provides the necessary ingredients for leptogenesis through the lepton-number-violating decays of heavy Majorana neutrinos. After diagonalizing the induced right-handed-neutrino Majorana mass matrix, the leptogenesis setup becomes close to the ordinary type-I seesaw case, with the heavy Majorana states $N_i$ decaying into lepton and Higgs final states. The decays proceed at tree and one-loop levels, as depicted in Fig.~\ref{fig:leptogenesis_diagram}, and the interference between these amplitudes generates a net CP asymmetry $\epsilon_{i\alpha}$, defined as \cite{Davidson:2008bu,Abada:2006fw,Nardi:2006fx}

\begin{equation}\label{CPialpha}
	\epsilon_{i\alpha}
	\equiv
	\frac{
		\Gamma\left(N_i \rightarrow  \ell_\alpha H \right)
		-
		\Gamma\left(N_i \rightarrow \bar{\ell}_\alpha H^* \right)
	}{
		\sum_\beta\left[
		\Gamma\left(N_i \rightarrow \ell_\beta H \right)
		+
		\Gamma\left(N_i \rightarrow \bar{\ell}_\beta H^* \right)
		\right]
	}
	\equiv
	\frac{\Delta \Gamma_{N_i}^\alpha}{\Gamma_{N_i}}.
\end{equation}
The departure from thermal equilibrium is characterized by comparing the decay rate $\Gamma_{N_i}$ with the Hubble expansion rate $H(T)$ around $T=M_i$. We define the decay parameter $K_i$ \cite{Davidson:2008bu,Giudice:2003jh} as
\begin{equation}\label{DecayParameter}
	K_i\equiv\frac{\Gamma_{N_i}}{H(M_i)}=\frac{\widetilde{m_i}}{m_*}
\end{equation}
where
\begin{eqnarray}\label{EffNuMassY}
	\begin{aligned}
		\widetilde{m_i}&\equiv\frac{(Y^\dagger Y)_{ii}v^2}{M_i}\text{ (effective neutrino mass)}\\
		m_*&\equiv\frac{16\pi^2 v^2\sqrt{g_*\pi/5}}{3M_{pl}}\approx 1.08\times10^{-3}eV
	\end{aligned}
\end{eqnarray}
The regimes $K_i \ll 1$, $K_i \approx 1$, and $K_i \gg 1$ are referred to as the weak, intermediate, and strong washout regimes, respectively. It should be noted that successful leptogenesis is not restricted only to the weak washout regime. In the strong washout regime, inverse decays initially keep the system close to equilibrium and erase any pre-existing asymmetry. However, the final asymmetry can still be generated when the washout processes become inefficient at later times.

In contrast to conventional type-I seesaw implementations, the right-handed neutrino masses in the present framework are not treated as arbitrary input parameters. They are generated through the double seesaw mechanism and are correlated with the light-neutrino sector via the non-holomorphic modular $A_4$ construction.

A further distinction appears when compared with our earlier double seesaw leptogenesis study in the left-right symmetric framework \cite{Patel:2023voj}. In that case, the presence of right-handed gauge interactions, especially the $W_R$-mediated processes, required additional scattering and washout effects to be included. In the present modular $A_4$ realization, no such extra right-handed gauge sector is present, so the leptogenesis dynamics is closer to the minimal type-I seesaw case. This makes the present setup technically simpler: for the unflavored benchmark discussed in this section, the observed baryon asymmetry is obtained already from the minimal decay and inverse-decay dynamics.

Having established the parameter space compatible with neutrino oscillation data in the previous section, we now proceed to the leptogenesis analysis. We consider viable normal-ordering solutions that also satisfy the adopted cosmological bound on the sum of light neutrino masses. Each allowed point is associated with a definite value of the lightest active neutrino mass $m_1$ and a corresponding RHN mass spectrum. The distribution of the resulting heavy neutrino masses is summarized in Fig.~\ref{fig:RHNspectrum} of Appendix~\ref{app:scan_diagnostics}, where the allowed points are shown in the $\left(M_{N_1},M_{N_2}\right)$ plane, with $M_{N_3}$ indicated by the color coding.

Rather than discussing the complete set of allowed points individually, we select representative benchmarks to study the different leptogenesis regimes realized in this framework. We first consider unflavored thermal leptogenesis, then the flavored regime, and finally examine whether the allowed RHN spectra also include quasi-degenerate configurations that can enter the resonance-sensitive regime. This possibility is explored through the scan-level diagnostic shown in Fig.~\ref{fig:quasidegenerate_diagnostic} of Appendix~\ref{app:scan_diagnostics}, where the $N_2$-$N_3$ mass splitting is compared with their average decay width while simultaneously monitoring the separation of $N_1$ from the quasi-degenerate pair and the associated RHN mass scale.

\subsection{Framework and relevant parameters}

Before discussing the different leptogenesis regimes through representative benchmark cases, we first collect the relevant parameters that enter the leptogenesis analysis in the present framework. This also fixes the notation used in the unflavored, flavored, and resonant discussions below.\\

\noindent\textbf{\underline{Physical basis of right-handed neutrinos}}\\
The relevant Dirac mass term in the interaction Lagrangian connects the left-handed neutrino fields with the right-handed neutrino fields, $\nu_L-N_R$, in the flavor basis of neutral fermions:
\begin{equation}\label{FlavorLagMD}
	\mathcal{L}_{M_D} = \sum_{\alpha,\beta} \overline{\nu_{\alpha L}}[M_D]_{\alpha\beta}N_{\beta R}+ h.c.
\end{equation}
However, the RHNs decay in their physical, or mass, basis, as represented in the Feynman diagrams in Fig.~\ref{fig:leptogenesis_diagram}. From the previous discussion, we know that the induced Majorana mass matrix of RHNs is not diagonal in the flavor basis. Thus, we need to rotate to a basis in which the RHN states correspond to physical mass eigenstates. Even though the bare Majorana mass term for $N_{Ri}$ is absent in the model, the double seesaw structure induces an effective Majorana mass term for $N_{Ri}$:
\begin{equation*}
	\frac{1}{2}\overline{N^{c}_{Ri}}\hspace{-1cm}\underbrace{M_{N_{ij}}}_{\text{induced Majorana mass}}\hspace{-1cm} N_{Rj}.
\end{equation*}
Diagonalizing $M_N$ brings us to the mass, or physical, basis. This can be done through Takagi decomposition:
\begin{equation*}
	U_R^T M_N U_R = \hat{m}_N \equiv \text{diag}(M_{N_1},M_{N_2}, M_{N_3})
\end{equation*}
Here, $\hat{m}_N$ contains the mass eigenvalues of $M_N$, and $U_R$ is the unitary matrix that diagonalizes $M_N$. This gives the relation between the physical states, $N_R^\text{m}$, and the flavor states, $N_R$, as \cite{Joshipura:2001ui,Davidson:2008bu}
\begin{equation*}
	N_R^{\text{m}} = U_R^\dagger N_R.
\end{equation*}
In this rotated basis, the Dirac neutrino mass matrix appearing in Eq.~(\ref{FlavorLagMD}) is also transformed as
\begin{equation}\label{MDR}
	M_{DR}=M_D\,U_R.
\end{equation}

\noindent\textbf{\underline{CP asymmetry}}\\
In the physical basis, the RHNs decay into final states containing leptons and anti-leptons with different decay rates, as shown in Fig.~\ref{fig:leptogenesis_diagram}. This difference generates a net lepton asymmetry, which is parametrized by the flavored CP asymmetry $\epsilon_{i\alpha}$ defined in Eq.~(\ref{CPialpha}). For a hierarchical RHN spectrum, with $N_i$ decaying into the lepton flavor $\alpha$ ($\alpha\equiv e,\mu,\tau$), the non-resonant CP asymmetry at leading order is given by \cite{Abada:2006fw,Nardi:2006fx,Davidson:2008bu}
\begin{equation}\label{CPialphaGeneral}
	\begin{aligned}
		\epsilon_{i \alpha}
		& =
		\frac{1}{8 \pi v^2}
		\frac{1}{\left(M_{DR}^{\dagger} M_{DR}\right)_{i i}}
		\sum_{j \neq i}
		\operatorname{Im}\left[
		\left(M_{DR}^{\dagger} M_{DR}\right)_{i j}
		\left(M_{DR}\right)_{\alpha i}^{*}
		\left(M_{DR}\right)_{\alpha j}
		\right]
		f\left(\frac{M_j^2}{M_i^2}\right)
		\\
		& +
		\frac{1}{8 \pi v^2}
		\frac{1}{\left(M_{DR}^{\dagger} M_{DR}\right)_{i i}}
		\sum_{j \neq i}
		\operatorname{Im}\left[
		\left(M_{DR}^{\dagger} M_{DR}\right)_{j i}
		\left(M_{DR}\right)_{\alpha i}^{*}
		\left(M_{DR}\right)_{\alpha j}
		\right]
		\frac{M_i^2}{M_i^2-M_j^2}.
	\end{aligned}
\end{equation}
where $f$ is the loop function defined as
\begin{equation}\label{CPloopFun}
	f(x)=\sqrt{x}\left[\frac{1}{1-x}+1-(1+x) \ln \left(\frac{1+x}{x}\right)\right].
\end{equation}
\textbf{Note:} The expression for the CP asymmetry $\epsilon_{i \alpha}$ in Eq.~(\ref{CPialphaGeneral}) is valid for a non-resonant, non-quasi-degenerate RHN mass spectrum. In the resonant case, the expression has to be replaced by a properly regulated form, which we discuss later in this section.\\

\noindent\textbf{\underline{Asymmetry from the $S_L$ sector}}\\
Apart from the RHNs, the present framework also contains three heavy left-handed Majorana singlets $S_L$. In the double seesaw realization considered here, these states lie at a much higher scale than the RHNs. The thermal histories adopted for the leptogenesis benchmarks are therefore taken below the sterile neutrino mass scale, so that the $S_L$ states are not thermally populated. We consequently do not follow their decay dynamics and focus on the asymmetry generated by the RHN sector.

Equipped with these relevant ingredients, we next discuss the different leptogenesis scenarios using representative benchmark points selected from the phenomenologically viable parameter space. These benchmark points specify the RHN mass eigenvalues, the Dirac neutrino mass matrix in the RHN mass basis, $M_{DR}$, and the lightest active neutrino mass $m_1$.

%%%%%%%%%%%%%%%%%%%%%%%%%%%%%%%%%%%%%%%%
\subsection{Unflavored thermal leptogenesis}

\begin{table}[t]
	\centering
	\renewcommand{\arraystretch}{1.35}
	\begin{tabular}{|c|c|c|c|c|}
		\hline
		\multicolumn{5}{|c|}{\textbf{Benchmark for Unflavored Thermal Leptogenesis}} \\
		\hline
		$m_1$ & $M_{N_1}$ & $M_{N_2}$ & $M_{N_3}$ & complex modulus $\tau$ \\
		\hline
		\begin{tabular}{c}
			$1.82\times 10^{-2}$ \\
			$(\mathrm{eV})$
		\end{tabular}
		&
		\begin{tabular}{c}
			$1.34\times 10^{13}$ \\
			$(\mathrm{GeV})$
		\end{tabular}
		&
		\begin{tabular}{c}
			$4.87\times 10^{13}$ \\
			$(\mathrm{GeV})$
		\end{tabular}
		&
		\begin{tabular}{c}
			$5.97\times 10^{13}$ \\
			$(\mathrm{GeV})$
		\end{tabular}
		&
		$-0.03+1.38\,i$
		\\
		\hline
	\end{tabular}
	\caption{Representative benchmark point used for thermal unflavored leptogenesis.}
	\label{tab:thermal-unflavored-benchmark}
\end{table}

In this subsection, we discuss a representative viable benchmark point of the model that realizes unflavored thermal leptogenesis. Table~\ref{tab:thermal-unflavored-benchmark} shows the benchmark inputs: the lightest active neutrino mass $m_1$, the RHN mass spectrum, and the complex modulus $\tau$ that controls the flavor structure and CP violation in the model. The Dirac neutrino mass matrix in the physical basis of RHNs, defined in Eq.~(\ref{MDR}), is

\begin{equation}
	{
		\setlength{\arraycolsep}{10pt}
		M_{DR}
		=	
		\begin{pmatrix}
			0.01+15.21\, i &
			0.07-2.75\, i &
			2.58+0.10\, i
			\\[1mm]
			0.10-0.71\, i &
			0.21+38.05\, i &
			-20.90+1.28\, i
			\\[1mm]
			0.57-11.87\, i &
			0.71-30.72\, i &
			-31.88-0.69\, i
		\end{pmatrix}\ {\rm GeV}.
	}
\end{equation}
Since the lightest RHN mass satisfies 
\begin{equation}
	M_{N_1} \simeq 1.34\times 10^{13}\ {\rm GeV} > 10^{12}\ {\rm GeV},
\end{equation}
the charged-lepton Yukawa interactions are out of equilibrium, and the evolution is described by the one-flavor or unflavored regime \cite{Abada:2006fw,Nardi:2006fx,Davidson:2008bu}. In the following analysis, we work in the $N_1$-dominated approximation. This can be realized if the reheating temperature after inflation satisfies $M_{N_1}\lesssim T_{\rm RH}\ll M_{N_2},M_{N_3}$, so that $N_1$ is thermally produced while the heavier states are not efficiently populated \cite{Fong:2012buy}.

For $N_1$ leptogenesis in the single flavor regime, we sum over the flavor index $\alpha$ in Eq.~(\ref{CPialphaGeneral}) with $i=1$ and obtain \cite{Giudice:2003jh,Davidson:2008bu}
\begin{equation}\label{epsilon1}
	\epsilon_1 \equiv \sum_\alpha \epsilon_{1 \alpha}=	\frac{1}{8\pi v^2}
	\frac{1}{[M_{DR}^\dagger M_{DR}]_{11}}
	\sum_{j\neq 1}
	{\rm Im}\left\{\left[(M_{DR}^\dagger M_{DR})_{1j}\right]^2\right\}
	f\left(\frac{M_j^2}{M_1^2}\right).
\end{equation}
For compactness, it is convenient to define 
\begin{equation}
	\mathcal{H}_D \equiv M_{DR}^{\dagger}M_{DR}.
\end{equation}
such that the unflavored CP asymmetry generated by $N_1$ decay in Eq.~(\ref{epsilon1}) is re-expressed as
\begin{equation}
	\epsilon_1
	=
\frac{1}{8\pi v^2}
\frac{1}{(\mathcal{H}_D)_{11}}
\sum_{j\neq 1}
{\rm Im}\!\left[((\mathcal{H}_D)_{1j})^2\right]\,
f\!\left(\frac{M_j^2}{M_1^2}\right),
	\label{eq:eps1-unflavored}
\end{equation}
where $f(x)$ is the loop-function defined in Eq.~(\ref{CPloopFun}), and $v\simeq 174$ GeV\footnote{$v\equiv v_H/\sqrt2\simeq174$ GeV as defined in Eq.~(\ref{scalarVEV}).}. The effective neutrino mass parameter and the corresponding decay parameter (defined in Eq.~(\ref{DecayParameter}) and Eq.~(\ref{EffNuMassY})) in terms of $\mathcal{H}_D$ are \cite{Giudice:2003jh,Davidson:2008bu}
\begin{equation}
	\widetilde m_1=\frac{(\mathcal{H}_D)_{11}}{M_1},
	\qquad
	K_1=\frac{\widetilde m_1}{m_*},
	\qquad
	m_*\simeq 1.08\times 10^{-3}\ {\rm eV}.
	\label{eq:tilde-m1-K1}
\end{equation}
For the present benchmark, we obtain
\begin{equation}
	\epsilon_1 \simeq -6.75\times 10^{-6},
	\qquad
	\widetilde m_1 \simeq 2.78\times 10^{-2}\ {\rm eV},
	\qquad
	K_1 \simeq 25.73.
\end{equation}

The value $K_1 \gg 1$ places the present benchmark deep in the strong-washout regime. The $N_1$-dominated thermal history is ensured by the reheating-temperature assumption discussed above, while the strong washout makes the final asymmetry largely insensitive to the initial $N_1$ abundance. In what follows, we first provide a semi-analytic estimate of the baryon asymmetry and then solve the Boltzmann equations to obtain the final asymmetry more concretely.\\

\noindent\textbf{\underline{Semi-analytic BAU estimate}}\\
For leptogenesis through the decays of $N_1$ in the one-flavor regime, the final baryon asymmetry can be estimated as \cite{Giudice:2003jh,Davidson:2008bu,Fong:2012buy}
\begin{equation}\label{YBest}
	Y_{\Delta B}^{\rm est}
	\simeq
	-\,10^{-3}\,
	\epsilon_1\,\eta_1\,C_{\rm sph}.
\end{equation}
Here, the factor $10^{-3}$ should be understood as the usual order-of-magnitude abundance and entropy-dilution factor appearing in the semi-analytic estimate. The minus sign reflects the convention that the generated $B-L$ asymmetry is opposite in sign to the lepton asymmetry produced in $N_1$ decays. The factor $\eta_1$ denotes the efficiency of the generated asymmetry after washout effects. In the strong-washout regime, it can be roughly approximated as
\begin{equation}
	\eta_1 \simeq \frac{1}{K_1}.
\end{equation}
The factor $C_{\rm sph}$ is the sphaleron conversion factor, which determines how much baryon asymmetry is produced from the $B-L$ asymmetry while electroweak sphalerons are active \cite{Harvey:1990qw}:
\begin{equation}
	C_{\rm sph}= \frac{28}{79}\quad(\text{SM}).
\end{equation}
Thus, for the benchmark point under consideration, the BAU can roughly be estimated as
\begin{equation}\label{YBestThermal}
	Y_{\Delta B}^{\rm est}
	\simeq
	-\,10^{-3}\times
	\left(-6.75\times 10^{-6}\right)
	\times
	\frac{1}{25.73}
	\times
	\frac{28}{79}
	\simeq
	9.30\times 10^{-11},
\end{equation}
which is already very close to the observed value
\begin{equation}
	Y_{\Delta B}^{\rm obs}\simeq 8.7\times 10^{-11}.
\end{equation}
The estimate in Eq.~(\ref{YBestThermal}) should only be regarded as an order-of-magnitude guide. We now solve the Boltzmann equations numerically to obtain the final baryon asymmetry more reliably.\\

\noindent\textbf{\underline{Boltzmann equations}}\\
To study the evolution of particle abundances, we use the yield variable, defined as the ratio of number density to entropy density:
\begin{equation}\label{AbundanceYi}
	Y_i\equiv \frac{n_i}{s},
\end{equation}
We solve the coupled Boltzmann equations for the lightest heavy neutrino yield $Y_{N_1}$ and the $B-L$ asymmetry $Y_{\Delta_{B-L}}$, considering only decays and inverse decays of $N_1$ \cite{Giudice:2003jh,Davidson:2008bu}:
\begin{align}
	\frac{dY_{N_1}}{dz}
	&=
	-\,D_1(z)\,
	\Bigl(Y_{N_1}-Y_{N_1}^{\rm eq}\Bigr),
\label{BEunflavored1}
	\\[1mm]
	\frac{dY_{\Delta_{B-L}}}{dz}
	&=
	-\epsilon_1\,D_1(z)\,
	\Bigl(Y_{N_1}-Y_{N_1}^{\rm eq}\Bigr)
	-
	W_1^{\rm ID}(z)\,Y_{\Delta_{B-L}}.
	\label{BEunflavored2}
\end{align}
where $z\equiv \frac{M_1}{T}$. The equilibrium heavy neutrino yield is
\begin{equation}
	Y_{N_1}^{\rm eq}(z)
	=	
	\frac{45}{2\pi^4 g_*}
	z^2 \mathcal{K}_2(z),
	\label{eq:YN1eq}
\end{equation}
with $g_*=106.75$ for the SM thermal bath, and the inverse-decay washout term is written as
\begin{equation}
	W_1^{\rm ID}(z)
	=	
	\frac{1}{2}D_1(z)
	\frac{Y_{N_1}^{\rm eq}(z)}{Y_\ell^{\rm eq}},
	\qquad
	Y_\ell^{\rm eq}
	=	
	\frac{15}{4\pi^2 g_*}.
	\label{eq:WID}
\end{equation}
The decay term takes the standard form
\begin{equation}
	D_1(z)
	=	
	K_1 z
	\frac{\mathcal{K}_1(z)}{\mathcal{K}_2(z)},
	\label{eq:D1}
\end{equation}
where $\mathcal{K}_n(z)$ denotes the $n$th order modified Bessel function of second kind.

We now solve the Boltzmann equations (Eq.~(\ref{BEunflavored1}) and Eq.~(\ref{BEunflavored2}))
for two choices of initial condition:
\begin{equation}
	Y_{N_1}(z_{\rm min})=Y_{N_1}^{\rm eq}(z_{\rm min})
	\qquad \text{(thermal initial abundance)},
\end{equation}
and
\begin{equation}
	Y_{N_1}(z_{\rm min})=0
	\qquad \text{(zero initial abundance)},
\end{equation}
with $z_{\rm min}=10^{-2}$ ($\equiv z\ll 1$) and $z_{\rm max}=100$ ($\equiv z\gg 1$). Before the electroweak phase transition, sphaleron processes partially convert the generated $B-L$ asymmetry into baryon asymmetry according to
\begin{equation}
	Y_{\Delta B}(z) = \frac{28}{79}Y_{\Delta_{B-L}}(z).
\end{equation}
For the present benchmark, the numerical solution gives
\begin{equation}
	Y_{\Delta B}^{\rm thermal}(z_{\rm max})
	\approx
	Y_{\Delta B}^{\rm zero}(z_{\rm max})
	\approx	
	8.11\times 10^{-11},
\end{equation}
with
\begin{equation}
	\left|
	Y_{\Delta B}^{\rm thermal}(z_{\rm max})-Y_{\Delta B}^{\rm zero}(z_{\rm max})
	\right|
	\simeq 1.6\times 10^{-23}.
\end{equation}
Therefore, the final asymmetry is essentially independent of the initial abundance. This is precisely what one expects in a strong-washout regime. We thus obtain
\begin{equation}
	Y_{\Delta B}^{\rm final}
	\simeq
	8.11\times 10^{-11}
	\simeq
	0.93\,Y_{\Delta B}^{\rm obs}.
\end{equation}
Hence, the benchmark successfully reproduces the observed baryon asymmetry already with decays and inverse decays alone.
\begin{figure}[t]
	\centering
	\begin{subfigure}[b]{0.60\textwidth}
		\includegraphics[width=\linewidth]{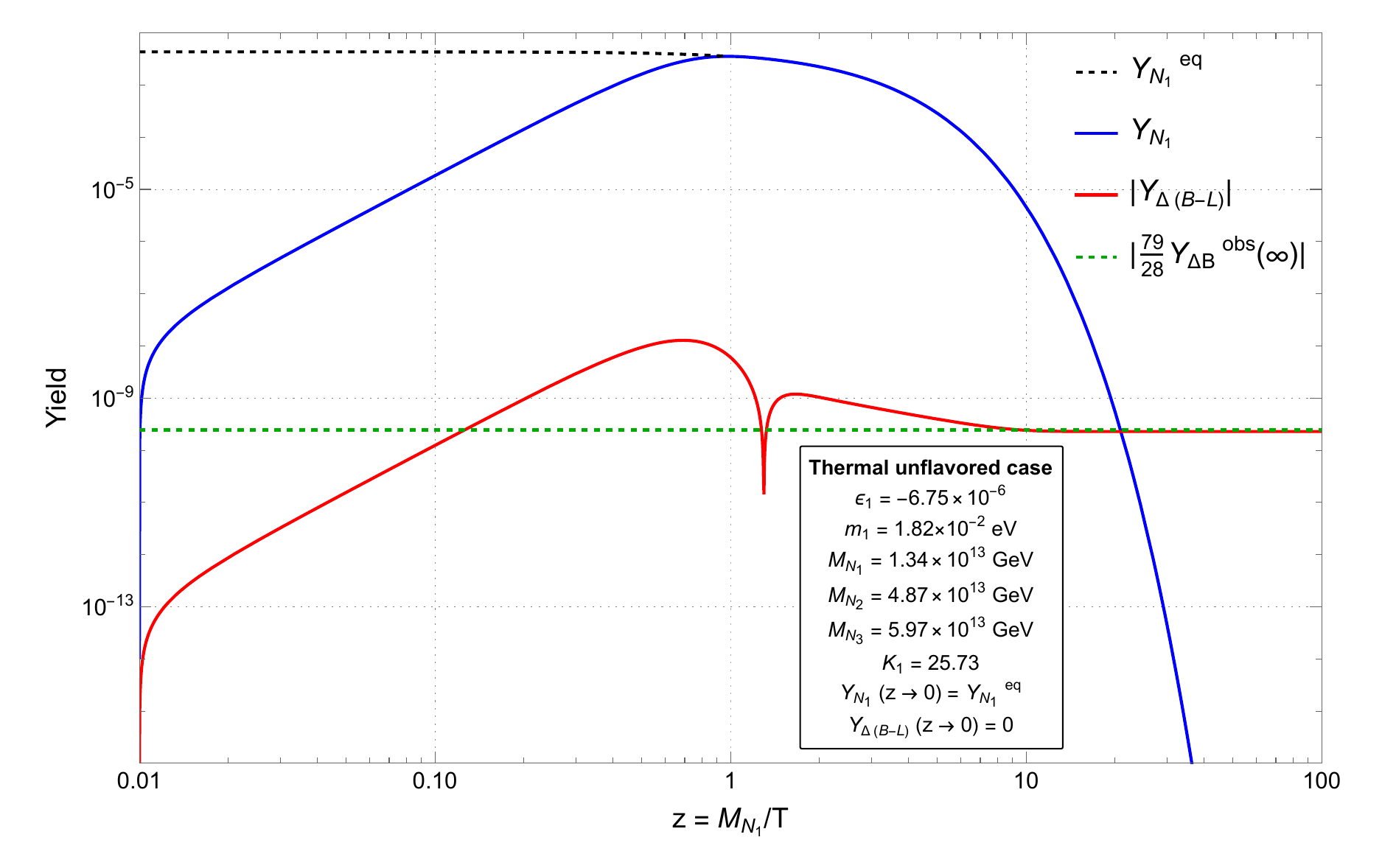}
		\caption{}
		\label{fig:unflavoredYield(a)} 
	\end{subfigure}
	%\hspace{0.0001in}
	\begin{subfigure}[b]{0.60\textwidth}
		\includegraphics[width=\linewidth]{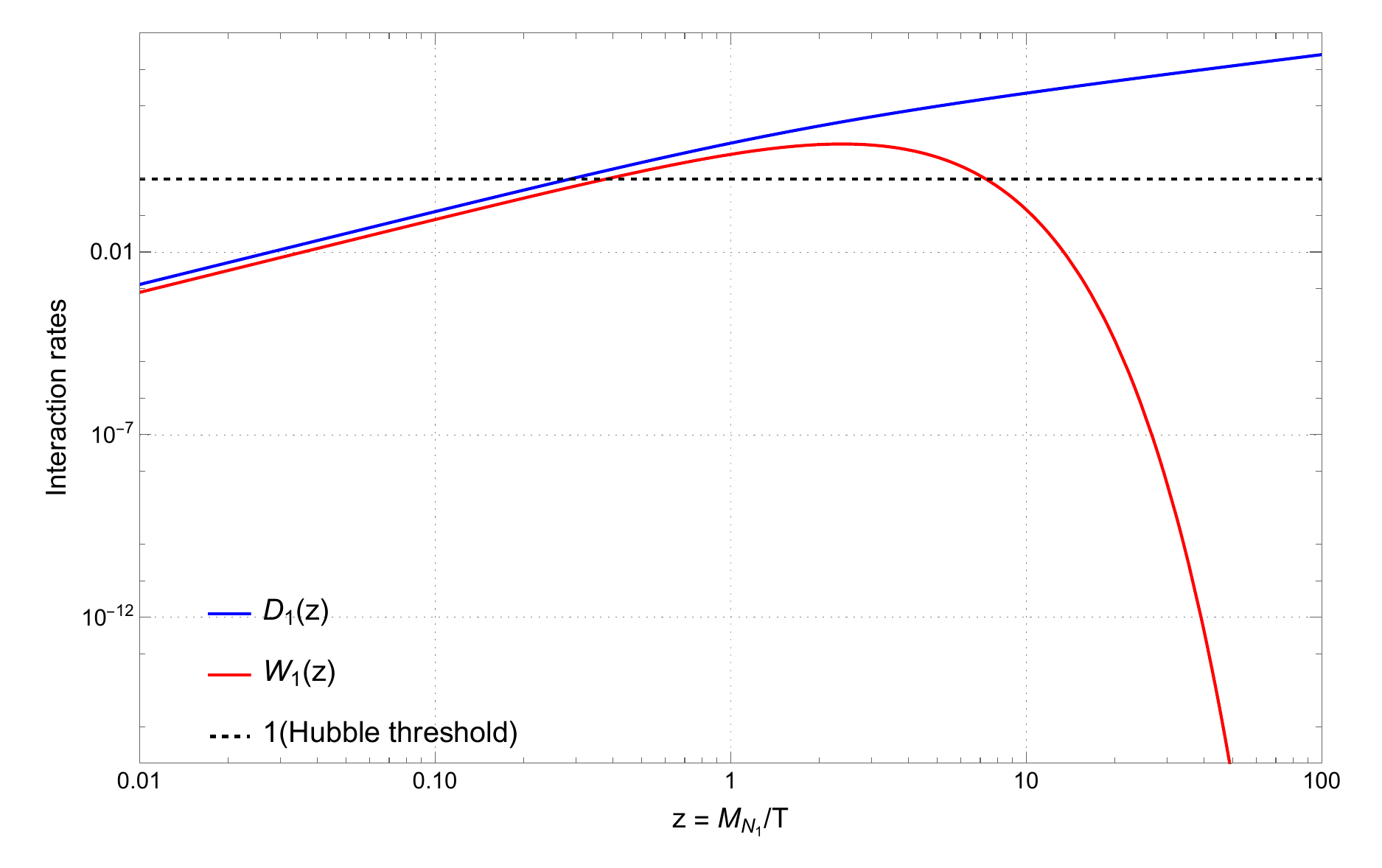}
		\caption{}
		\label{fig:unflavoredRates(b)}
	\end{subfigure}
	\caption{Panel (a) shows the cosmological evolution of the relevant yields for the unflavored thermal benchmark: the heavy neutrino yield $Y_{N_1}$, the equilibrium yield $Y_{N_1}^{\rm eq}$, and the generated asymmetry $|Y_{\Delta_{B-L}}|$ as functions of $z=M_1/T$. The horizontal green dashed line corresponds to the observed baryon asymmetry translated to the $B-L$ yield. The plot exhibits the standard strong-washout behaviour: $Y_{N_1}$ rapidly approaches equilibrium, tracks it over a broad range of $z$, and the asymmetry freezes to a value close to the observed one once inverse decays decouple. Panel (b) shows the interaction-rate plot for the same benchmark. The decay term $D_1(z)$ and inverse-decay washout term $W_1^{\rm ID}(z)$ are plotted together with the reference line $W=1$. The two crossings of $W_1^{\rm ID}$ with unity determine the wash-in and freeze-out points, approximately at $z\simeq 0.38$ and $z\simeq 7.30$, respectively. This confirms that the benchmark lies in the strong-washout regime and explains why the final baryon asymmetry is essentially independent of the initial $N_1$ abundance.}
\end{figure}

The numerical solution of the Boltzmann equations for the benchmark point is shown in Fig.~\ref{fig:unflavoredYield(a)}. The plot displays the yields $Y_{N_1}$, $Y_{N_1}^{\rm eq}$, and $|Y_{\Delta_{B-L}}|$ as functions of $z=M_{N_1}/T$. At very early times, $z\ll 1$, the temperature of the Universe is high, and $N_1$ can be thermally produced. The solid blue curve shows the cosmological evolution of the $N_1$ abundance. It rapidly approaches the equilibrium abundance $Y_{N_1}^{\rm eq}$, shown by the black dashed curve, and tracks it over a broad range of $z$. At later times, once $N_1$ becomes non-relativistic, its abundance decreases as expected. The red solid curve shows the evolution of the $B-L$ asymmetry. For visual clarity, we show the absolute value $|Y_{\Delta_{B-L}}|$. To compare with the observed baryon asymmetry, we translate it to the corresponding $B-L$ value through $|Y_{\Delta_{B-L}}^{\rm obs}|=\frac{79}{28}|Y_{\Delta B}^{\rm obs}|$, shown by the green dashed line. Since the thermal and zero-initial-abundance solutions are indistinguishable at the level of the final asymmetry, the plotted branch may be taken as representative of either. The $B-L$ asymmetry grows mainly once the departure from equilibrium becomes appreciable and then freezes to a constant value, very close to the observed $B-L$ asymmetry, after the washout term becomes inefficient.

To complement the yield plot in Fig.~\ref{fig:unflavoredYield(a)}, we also show the variation of interaction rates with respect to $z$ in Fig.~\ref{fig:unflavoredRates(b)}. The decay term $D_1(z)$ and the inverse-decay washout term $W_1^{\rm ID}(z)$ are plotted together with the reference line $W=1$. Since these quantities are normalized to the Hubble expansion rate in the Boltzmann equations, the line $W=1$ acts as the Hubble threshold for washout efficiency \cite{Giudice:2003jh}. The first crossing of $W_1^{\rm ID}(z)$ with unity marks the onset of efficient inverse-decay washout, while the second crossing marks the freeze-out of this washout. In the interval between these two crossings,
\begin{equation}
	0.38 \lesssim z \lesssim 7.30,
\end{equation}
inverse decays are faster than the Hubble expansion and efficiently erase any generated asymmetry. This explains both the loss of sensitivity to the initial condition and the smooth saturation of the final asymmetry. After the second crossing, inverse decays become Boltzmann suppressed, and the generated asymmetry freezes to its final value.

At $z=1$ we find
\begin{equation}
	D_1(1)\simeq 9.53,
	\qquad
	W_1^{\rm ID}(1)\simeq 4.71.
\end{equation}
This is the characteristic epoch when $N_1$ transitions from relativistic to non-relativistic behaviour. Since both $D_1(1)$ and $W_1^{\rm ID}(1)$ are much larger than $1$, the decay/inverse-decay processes are faster than the Hubble expansion. Therefore, $N_1$ is efficiently produced and remains close to its equilibrium abundance. Hence, the system is in the regime of efficient production and washout.

Before moving to the second benchmark point for the flavored leptogenesis study, it is worth emphasizing that the present benchmark reproduces the observed BAU within the minimal decay and inverse-decay system. No additional scattering terms are required to obtain the correct order of baryon asymmetry for this representative point.

%%%%%%%%%%%%%%%%%%%%%%%%%%%%%%%%%%%%%%%%%%
\subsection{$N_1$-dominated two-flavor thermal leptogenesis}

Although the unflavored benchmark already demonstrates that the observed baryon asymmetry can be reproduced in the present framework, it is useful to examine whether the model also admits a viable realization in which charged-lepton flavor effects become relevant. For temperatures below approximately $10^{12}\,{\rm GeV}$, the interactions mediated by the $\tau$ Yukawa coupling can distinguish the $\tau$ component of the lepton state produced in RHN decays, while the $e$ and $\mu$ components remain coherent down to temperatures of order $10^9\,{\rm GeV}$ \cite{Abada:2006ea,Davidson:2008bu}. The corresponding intermediate regime is therefore described by two effective flavor directions, conventionally denoted by $\tau$ and $\gamma\equiv e+\mu$. We identify a representative oscillation-compatible parameter point for which $M_{N_1}$ lies within this two-flavor window and use it to examine flavored thermal leptogenesis through a semi-analytic treatment.

\begin{table}[t]
	\centering
	\renewcommand{\arraystretch}{1.35}
	\begin{tabular}{|c|c|c|c|c|}
		\hline
		\multicolumn{5}{|c|}{\textbf{Benchmark for Two-Flavor Thermal Leptogenesis}} \\
		\hline
		$m_1$ & $M_{N_1}$ & $M_{N_2}$ & $M_{N_3}$ & complex modulus $\tau$ \\
		\hline
		\begin{tabular}{c}
			$3.27\times 10^{-3}$ \\
			$(\mathrm{eV})$
		\end{tabular}
		&
		\begin{tabular}{c}
			$1.82\times 10^{11}$ \\
			$(\mathrm{GeV})$
		\end{tabular}
		&
		\begin{tabular}{c}
			$1.16\times 10^{12}$ \\
			$(\mathrm{GeV})$
		\end{tabular}
		&
		\begin{tabular}{c}
			$1.78\times 10^{12}$ \\
			$(\mathrm{GeV})$
		\end{tabular}
		&
		$0.398+0.998\,i$
		\\
		\hline
	\end{tabular}
	\caption{Representative benchmark point used for the $N_1$-dominated two-flavor thermal leptogenesis analysis.}
	\label{tab:thermal-flavored-benchmark}
\end{table}

For this benchmark, the RHN mass spectrum, the lightest active neutrino mass $m_1$, and the corresponding complex modulus $\tau$ are summarized in Table~\ref{tab:thermal-flavored-benchmark}. The rotated Dirac neutrino mass matrix in the physical RHN basis is
\begin{equation}
	\begingroup
	\setlength{\arraycolsep}{3pt}
	\begin{aligned}
		&\frac{M_{DR}}{{\rm GeV}}=
		\\[1mm]
		&
		\begin{pmatrix}
			(-3.99-1.75\,i)\times10^{-1}
			&
			(2.01-5.34\,i)\times10^{-1}
			&
			2.17\times10^{-1}+2.17\times10^{-2}\,i
			\\[2mm]
			2.32+1.88\,i
			&
			(-3.06-3.67\,i)\times10^{-1}
			&
			1.93+5.76\times10^{-1}\,i
			\\[2mm]
			1.05+1.82\,i
			&
			5.42+2.16\,i
			&
			-4.64-6.79\,i
		\end{pmatrix}.
	\end{aligned}
	\endgroup
\end{equation}
The lightest RHN mass,
\begin{equation}
	M_{N_1}=1.82\times10^{11}\ {\rm GeV},
\end{equation}
lies between $10^9$ and $10^{12}\,{\rm GeV}$. At this scale the $\tau$ flavor is resolved, whereas the $e$ and $\mu$ components remain coherent and must be treated as the single coherent direction $\gamma\equiv e+\mu$ \cite{Abada:2006ea,Davidson:2008bu}.\footnote{We have verified that the charged-lepton flavor-decoherence condition remains satisfied despite the strong inverse-decay washout for this benchmark, ensuring the validity of the classical two-flavor treatment \cite{Blanchet:2008pw}.}

We consider an $N_1$-dominated thermal history in which the reheating temperature is sufficiently high to thermally populate $N_1$ but remains below the heavier RHN scales. The contributions from thermally produced $N_2$ and $N_3$ states can therefore be neglected, so that both the generation of the lepton asymmetry and its inverse-decay washout are governed by the $N_1$ dynamics and are encoded through the corresponding flavor-dependent efficiency factors.

The selected benchmark also exhibits a sufficiently hierarchical RHN mass spectrum,
\begin{equation}
	\frac{M_{N_2}}{M_{N_1}}\simeq6.38,
	\qquad
	\frac{M_{N_3}}{M_{N_1}}\simeq9.74.
\end{equation}
This hierarchy ensures that the flavored asymmetry can be treated within the standard non-resonant framework. To verify explicitly that no resonant enhancement is involved, we compare the pairwise RHN mass splittings with their corresponding average decay widths and obtain
\begin{equation}
	\min_{i<j}
	\left(
	\frac{\Delta M_{ij}}{\Gamma_{ij}^{\rm avg}}
	\right)
	\simeq5.56\times10^3\gg1.
\end{equation}
The benchmark therefore lies well outside the resonant regime, and the baryon asymmetry discussed below originates from ordinary flavored thermal leptogenesis.

In the two-flavor regime, the relevant flavored CP asymmetries are those associated with the resolved $\tau$ direction and the coherent $\gamma=e+\mu$ direction. They are defined as \cite{Abada:2006ea,Davidson:2008bu}
\begin{equation}
	\epsilon_{1\tau}
	\quad \text{and}\quad
	\epsilon_{1\gamma}
	=
	\epsilon_{1e}+\epsilon_{1\mu},
\end{equation}
where the individual $\epsilon_{1\alpha}$ are evaluated from the general flavored CP-asymmetry expression given in Eq.~(\ref{CPialphaGeneral}). For the present benchmark, we obtain
\begin{equation}
	\epsilon_{1e}\simeq-4.56\times10^{-8},
	\qquad
	\epsilon_{1\mu}\simeq-5.28\times10^{-7},
	\qquad
	\epsilon_{1\tau}\simeq3.16\times10^{-6},
\end{equation}
and consequently
\begin{equation}
	\epsilon_{1\gamma}
	\simeq-5.74\times10^{-7},
	\qquad
	\epsilon_{1\tau}
	\simeq3.16\times10^{-6}.
\end{equation}
The inverse-decay washout along each flavor direction is characterized by the effective neutrino mass and the corresponding decay parameter. For an individual charged-lepton flavor,
\begin{equation}
	\widetilde m_{1\alpha}
	=
	\frac{|(M_{DR})_{\alpha1}|^2}{M_{N_1}},
	\qquad
	K_{1\alpha}
	=
	\frac{\widetilde m_{1\alpha}}{m_*},
	\qquad
	m_*\simeq1.08\times10^{-3}\ {\rm eV}.
\end{equation}
Since $e$ and $\mu$ are not independently resolved at $T\sim M_{N_1}$, the physical washout parameter in the coherent non-$\tau$ direction is instead
\begin{equation}
	\widetilde m_{1\gamma}
	=
	\widetilde m_{1e}+\widetilde m_{1\mu},
	\qquad
	K_{1\gamma}
	=
	K_{1e}+K_{1\mu}.
\end{equation}
Numerically, we find
\begin{equation}
	\widetilde m_{1\gamma}
	\simeq4.995\times10^{-2}\ {\rm eV},
	\qquad
	\widetilde m_{1\tau}
	\simeq2.423\times10^{-2}\ {\rm eV},
\end{equation}
which give
\begin{equation}
	K_{1\gamma}\simeq46.25,
	\qquad
	K_{1\tau}\simeq22.44.
\end{equation}
Both resolved flavor directions therefore lie firmly in the strong-washout regime. This is particularly convenient for the present semi-analytic treatment, since the final asymmetry becomes insensitive to the weak-washout behavior of the efficiency factors.

To estimate the final asymmetry, we follow the standard two-flavor semi-analytic treatment of Ref.~\cite{Abada:2006ea}. In the thermal plasma, fast Standard Model spectator processes redistribute the generated asymmetries among the different particle species. It is therefore convenient to describe the flavor asymmetries in terms of the charges $\Delta_\alpha\equiv B/3-L_\alpha$, which are conserved by the Standard Model interactions relevant during leptogenesis and are violated by the lepton-number-violating RHN interactions. In the two-flavor regime, the corresponding independent directions are $\gamma=e+\mu$ and $\tau$.

The spectator effects modify the relation between the asymmetries stored in the lepton doublets and the corresponding $\Delta_\alpha$ charges. Following the diagonal approximation of Ref.~\cite{Abada:2006ea}, these effects can be incorporated in the semi-analytic washout expressions through the replacements
\begin{equation}
	\widetilde m_{1\gamma}
	\rightarrow
	\frac{417}{589}\,
	\widetilde m_{1\gamma}\,
	\qquad
	\widetilde m_{1\tau}
	\rightarrow
	\frac{390}{589}\,
	\widetilde m_{1\tau}.
\end{equation}
For the present benchmark, the corresponding quantities entering the washout expressions are
\begin{equation}
	\frac{417}{589}\widetilde m_{1\gamma}
	\simeq3.54\times10^{-2}\ {\rm eV},
	\qquad
	\frac{390}{589}\widetilde m_{1\tau}
	\simeq1.60\times10^{-2}\ {\rm eV}.
\end{equation}

In the strong-washout regime, Ref.~\cite{Abada:2006ea} gives the asymmetry in each resolved lepton-flavor direction in the approximate form
\begin{equation}
	Y_{\alpha}
	\simeq
	0.3\,
	\frac{\epsilon_{1\alpha}}{g_*}
	\left(
	\frac{0.55\times10^{-3}\ {\rm eV}}
	{|A_{\alpha\alpha}|\,\widetilde m_{1\alpha}}
	\right)^{1.16},
	\qquad
	\alpha=\gamma,\tau,
\end{equation}
where
\begin{equation}
	|A_{\gamma\gamma}|=\frac{417}{589},
	\qquad
	|A_{\tau\tau}|=\frac{390}{589}.
\end{equation}
For convenience, we write
\begin{equation}
	Y_{\alpha}
	\equiv
	\frac{\epsilon_{1\alpha}}{g_*}\eta_{1\alpha},
\end{equation}
so that the corresponding flavor-dependent efficiencies are
\begin{equation}
	\eta_{1\gamma}
	=
	0.3
	\left[
	\frac{0.55\times10^{-3}\ {\rm eV}}
	{(417/589)\widetilde m_{1\gamma}}
	\right]^{1.16},
	\qquad
	\eta_{1\tau}
	=
	0.3
	\left[
	\frac{0.55\times10^{-3}\ {\rm eV}}
	{(390/589)\widetilde m_{1\tau}}
	\right]^{1.16}.
\end{equation}
Numerically, we obtain
\begin{equation}
	\eta_{1\gamma}
	\simeq2.40\times10^{-3},
	\qquad
	\eta_{1\tau}
	\simeq5.99\times10^{-3}.
\end{equation}
These efficiency factors account for the production of the $N_1$-generated asymmetry together with the corresponding inverse-decay washout in each resolved flavor direction.

Using the two-flavor baryon conversion relation of Ref.~\cite{Abada:2006ea}, the final baryon asymmetry can be written as\footnote{The factor $12/37$ follows the flavor-asymmetry convention employed in the two-flavor semi-analytic treatment of Ref.~\cite{Abada:2006ea} and should not be identified directly with the $28/79$ conversion applied to the total $B-L$ yield in the unflavored Boltzmann treatment above.}
\begin{equation}
	Y_{\Delta B}
	\simeq
	-\frac{12}{37}
	\left[
	\frac{\epsilon_{1\gamma}}{g_*}\eta_{1\gamma}
	+
	\frac{\epsilon_{1\tau}}{g_*}\eta_{1\tau}
	\right],
	\qquad
	g_*=106.75.
	\label{eq:YB_twoflavor}
\end{equation}
Here the explicit factor $1/g_*$ follows from the strong-washout flavor asymmetry appearing above. For the present benchmark, the two flavor contributions are
\begin{equation}
	\frac{\epsilon_{1\gamma}}{g_*}\eta_{1\gamma}
	\simeq
	-1.29\times10^{-11},
	\qquad
	\frac{\epsilon_{1\tau}}{g_*}\eta_{1\tau}
	\simeq
	1.78\times10^{-10}.
\end{equation}
The final asymmetry is therefore dominated by the $\tau$ direction, while the coherent $\gamma$ contribution partially compensates it. Taking the magnitude, we obtain
\begin{equation}
	|Y_{\Delta B}|_{\rm semi-an.}
	\simeq
	5.34\times10^{-11},
\end{equation}
corresponding to
\begin{equation}
	\frac{|Y_{\Delta B}|_{\rm semi-an.}}{Y_{\Delta B}^{\rm obs}}
	\simeq0.61,
	\qquad
	Y_{\Delta B}^{\rm obs}\simeq8.7\times10^{-11}.
\end{equation}
The semi-analytic estimate therefore yields a baryon asymmetry of the phenomenologically relevant magnitude. This indicates that the oscillation-compatible parameter space of the model can also support an $N_1$-dominated two-flavor leptogenesis realization with a hierarchical, non-resonant RHN spectrum. The flavor-dependent CP asymmetries and strong-washout efficiencies generate a sizable final baryon asymmetry without relying on resonant enhancement. A more precise assessment of this benchmark would require solving the full flavored Boltzmann equations, or more generally the corresponding kinetic equations including the coupled flavor evolution. Such a detailed numerical treatment is beyond the scope of the present work, and the benchmark is therefore used here to demonstrate the viability of flavored thermal leptogenesis within the non-holomorphic modular $A_4$ double seesaw framework.

%%%%%%%%%%%%%%%%%%%%%%%%%%%%%%%%%%%%%%%%%%%
\subsection{Diagnostic of the quasi-degenerate regime}

Having established viable hierarchical realizations of unflavored and flavored thermal leptogenesis, it is also useful to examine whether the RHN spectrum allowed by the model contains quasi-degenerate configurations that can enter the resonant regime. Such a possibility is particularly interesting in the present framework because the RHN masses are not introduced independently, but arise from the underlying double seesaw structure. We therefore investigate whether the allowed parameter space naturally contains pairs of RHNs whose mass splitting becomes comparable to their decay widths, which is the characteristic condition for resonantly enhanced self-energy effects \cite{Pilaftsis:1997dr,Pilaftsis:2003gt,Bambhaniya:2016rbb}.

For a pair of RHNs $N_i$ and $N_j$, we define
\begin{equation}
	\mathcal{R}_{ij}
	\equiv
	\frac{\Delta M_{ij}}{\Gamma_{ij}^{\rm avg}}
	=
	\frac{|M_{N_j}-M_{N_i}|}
	{(\Gamma_i+\Gamma_j)/2}.
	\label{eq:Rij_def}
\end{equation}
The resonance-sensitive region corresponds to $\mathcal{R}_{ij}\sim\mathcal{O}(1)$. In the scan, the quasi-degenerate structure is most conveniently examined for the heavier $(N_2,N_3)$ pair through $\mathcal{R}_{23}$. To determine whether this pair is isolated from the lightest RHN, we additionally introduce
\begin{equation}
	\mathcal{R}_{1(23)}
	\equiv
	\min\left(\mathcal{R}_{12},\mathcal{R}_{13}\right).
	\label{eq:R1pair_def}
\end{equation}
A value $\mathcal{R}_{23}\sim1$ therefore indicates a resonance-sensitive $(N_2,N_3)$ pair, while $\mathcal{R}_{1(23)}\gg1$ shows that $N_1$ remains well separated from both heavier states.

The corresponding scan diagnostic is shown in Fig.~\ref{fig:quasidegenerate_diagnostic} of Appendix~\ref{app:scan_diagnostics}. The horizontal axis displays $\log_{10}\mathcal{R}_{23}$, such that points near the center of the resonance-sensitive band satisfy $\Delta M_{23}\sim\Gamma_{23}^{\rm avg}$. The vertical axis shows $\log_{10}\mathcal{R}_{1(23)}$ and therefore measures the separation of $N_1$ from the heavier pair in units of the corresponding decay widths. Points lying near $\log_{10}\mathcal{R}_{23}\simeq0$ and simultaneously at large $\log_{10}\mathcal{R}_{1(23)}$ thus represent the cleanest configurations in which $N_2$ and $N_3$ form an isolated quasi-degenerate pair. The color scale indicates the average mass of this pair,
\begin{equation}
	M_{23}^{\rm avg}
	=
	\frac{M_{N_2}+M_{N_3}}{2},
\end{equation}
and therefore identifies the RHN mass scale at which such configurations occur.

The diagnostic shows that the oscillation-compatible parameter space can indeed contain RHN spectra approaching the resonance-sensitive condition for the $(N_2,N_3)$ pair. These configurations occur at comparatively high RHN mass scales, where conventional thermal leptogenesis can already generate a sizable baryon asymmetry without requiring resonant enhancement, as demonstrated by the hierarchical benchmarks discussed above. The quasi-degenerate points, therefore, indicate that the model also admits RHN spectra in which resonant effects may become relevant.

We emphasize that Fig.~\ref{fig:quasidegenerate_diagnostic} is intended only as a diagnostic of the RHN spectrum and not as a calculation of the resonant baryon asymmetry. Once $\Delta M_{23}$ becomes comparable to the decay widths, a quantitative prediction requires properly regulated resonant CP asymmetries together with a kinetic treatment that accounts for the mixing and possible coherence of the nearly degenerate RHNs \cite{Pilaftsis:1997dr,Pilaftsis:2003gt}. Such an analysis lies beyond the scope of the present work. Thus, the present study establishes successful leptogenesis in the non-resonant regime while also showing that resonance-sensitive quasi-degenerate RHN spectra occur within the allowed parameter space.

\section{Conclusions}\label{sec:conclusion}
In this work, we have developed a non-holomorphic modular $A_4$ realization of the double seesaw mechanism, in which the underlying seesaw architecture is augmented by modular symmetry in a way that goes beyond merely constraining the flavor textures. We have shown that the Majorana mass matrix of the right-handed neutrinos need not be introduced as an independent input, as in the canonical Type-I seesaw, since the modular weight assignments forbid a bare Majorana mass term and enforce its generation through the double seesaw structure. In this way, we realize the masses of the light active neutrinos, right-handed neutrinos, and sterile neutrinos within a unified framework governed by a common symmetry structure. We have further conducted a comprehensive analysis of the resulting parameter space and its phenomenological consequences in both the low- and high-energy neutrino sectors. We find that the model simultaneously accommodates the observed neutrino oscillation data, admits phenomenologically viable solutions only for NO in the active neutrino sector within the explored parameter space, and yields right-handed neutrino masses together with symmetry-constrained Yukawa textures that successfully reproduce the observed BAU through thermal leptogenesis. At the same time, the resulting oscillation predictions remain testable through current and forthcoming precision neutrino measurements. We therefore find that the proposed framework provides a common setting in which neutrino masses, oscillation phenomenology, and the cosmological matter-antimatter asymmetry emerge as interconnected consequences of the same underlying construction.

Numerically, we have explored the modular parameter space over $-0.5\leq\mathrm{Re}(\tau)\leq0.5$ and $0.86\leq\mathrm{Im}(\tau)\leq5$, subject to $|\tau|\geq1$, while varying the relevant dimensionless coupling ratios from $10^{-8}$ to $10^{4}$. Within the viable NO parameter space, the best-fit point is obtained at $\mathrm{Re}(\tau)=-0.018$ and $\mathrm{Im}(\tau)=1.65$. The model reproduces the measured mass-squared differences and mixing angles while constraining $\sin^2\theta_{13}$ to the narrow interval $[0.020,0.024]$, whereas the predicted $\delta_{\rm CP}$ spans its entire physical range. We have further confronted these predictions with the projected sensitivities of DUNE, T2HK, and JUNO. Under the lower-octant true hypothesis adopted in the long-baseline precision simulations, DUNE and T2HK substantially narrow the atmospheric parameter space and provide strong discrimination between the lower- and higher-octant regions allowed by the model, with the projected sensitivities providing a direct test of the higher-octant solutions. JUNO provides complementary sensitivity to $\theta_{12}$ and $\Delta m^2_{21}$, with its emerging data already constraining the solar neutrino parameter space. Looking ahead, the projected full-exposure sensitivity offers a precise probe of both the solar parameters and $\Delta m^2_{31}$. These current and projected measurements can therefore significantly restrict the presently allowed model parameter space and provide complementary tests of its low-energy predictions.

The predictive structure of the model also extends to the high-energy sector, since each oscillation-compatible parameter point determines the lightest active neutrino mass, the induced RHN spectrum, and the corresponding complex Dirac neutrino mass matrix required for leptogenesis. We have studied representative benchmark points that realize different thermal leptogenesis regimes through the out of equilibrium decays of the heavy Majorana neutrinos. For the unflavored benchmark, characterized by $m_1=1.82\times10^{-2}$ eV and $M_{N_1}=1.34\times10^{13}$ GeV, we obtain a CP asymmetry $\epsilon_1\simeq-6.75\times10^{-6}$ and a decay parameter $K_1\simeq25.73$, placing the system firmly in the strong-washout regime. Solving the coupled Boltzmann equations for $N_1$ decays and inverse decays gives $Y_{\Delta B}\simeq8.11\times10^{-11}\simeq0.93\,Y_{\Delta B}^{\rm obs}$. The thermal and zero initial-$N_1$ abundance solutions converge to essentially the same final asymmetry, demonstrating the expected loss of sensitivity to the initial conditions in the strong-washout regime and showing that the minimal decay and inverse-decay dynamics can already reproduce the observed BAU.

We have further identified an $N_1$-dominated two-flavor realization with $m_1=3.27\times10^{-3}$ eV and a hierarchical spectrum $M_{N_1}=1.82\times10^{11}$ GeV, $M_{N_2}=1.16\times10^{12}$ GeV, and $M_{N_3}=1.78\times10^{12}$ GeV. In this regime, the $\tau$ flavor is resolved while the $e$ and $\mu$ components remain coherent. The semi-analytic strong-washout treatment gives $|Y_{\Delta B}|\simeq5.34\times10^{-11}\simeq0.61\,Y_{\Delta B}^{\rm obs}$, demonstrating that the oscillation-compatible parameter space also supports flavored thermal leptogenesis at the phenomenologically relevant scale. A more precise assessment of this benchmark would require the full flavored kinetic evolution, which is beyond the scope of the present work. The allowed parameter space also contains points for which the heavier $(N_2,N_3)$ pair satisfies $\Delta M_{23}/\Gamma_{23}^{\rm avg}\sim\mathcal{O}(1)$ while $N_1$ remains well separated, showing that the framework can naturally accommodate resonance-sensitive RHN spectra. Since these resonance-sensitive configurations themselves occur at comparatively high RHN mass scales, quasi-degeneracy is not required to overcome the usual suppression of the CP asymmetry associated with low-scale leptogenesis. They are therefore best regarded as an additional feature of the RHN spectrum allowed by the framework, rather than as a necessary ingredient for explaining the observed BAU.

%%%%%%%%%%%%%%%%%%%%%%%%%%%%%%%%%%%%%%%%%%%%%%%%%%%%
\section*{Acknowledgements}
PA would like to acknowledge the financial support obtained from the Ministry of Education, Government of India. DKS would like to thank Dr.~Monojit Ghosh for useful discussions and would also like to acknowledge Ministry of Science and Education of Republic of Croatia grant No.~PK.1.1.10.0002, Swiss National Science Foundation (SNSF) and Croatian Science Foundation (HRZZ) under grant MAPS IZ11Z0$\_$230193 for financial support. SP acknowledges the Institute of Physics, Bhubaneswar, for hospitality during his sabbatical stay, where part of this work was carried out, and the funding support from SERB, Government of India, under the MATRICS project, Grant No.~MTR/2023/000687.

\newpage
\appendix

\section{Heavy neutrino spectra and resonance diagnostics}
\label{app:scan_diagnostics}
\begin{figure}[htbp]
	\centering
	\includegraphics[width=0.65\textwidth]{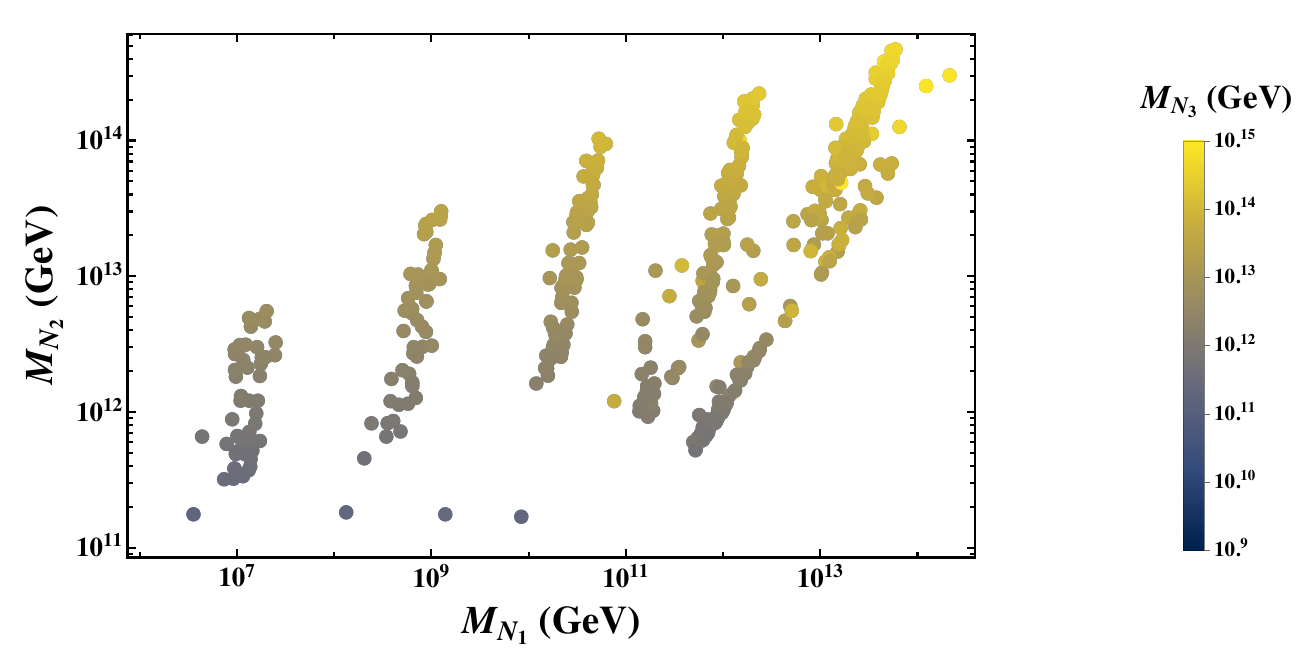}
	\caption{
		Distribution of RHN mass spectra obtained from the viable normal-ordering parameter sets. The horizontal and vertical axes show $M_{N_1}$ and $M_{N_2}$, respectively, while the color bar denotes $M_{N_3}$. Each point corresponds to a parameter set satisfying the low-energy neutrino oscillation constraints.
	}
	\label{fig:RHNspectrum}
\end{figure}

\begin{figure}[htbp]
	\centering
	\includegraphics[width=0.65\textwidth]{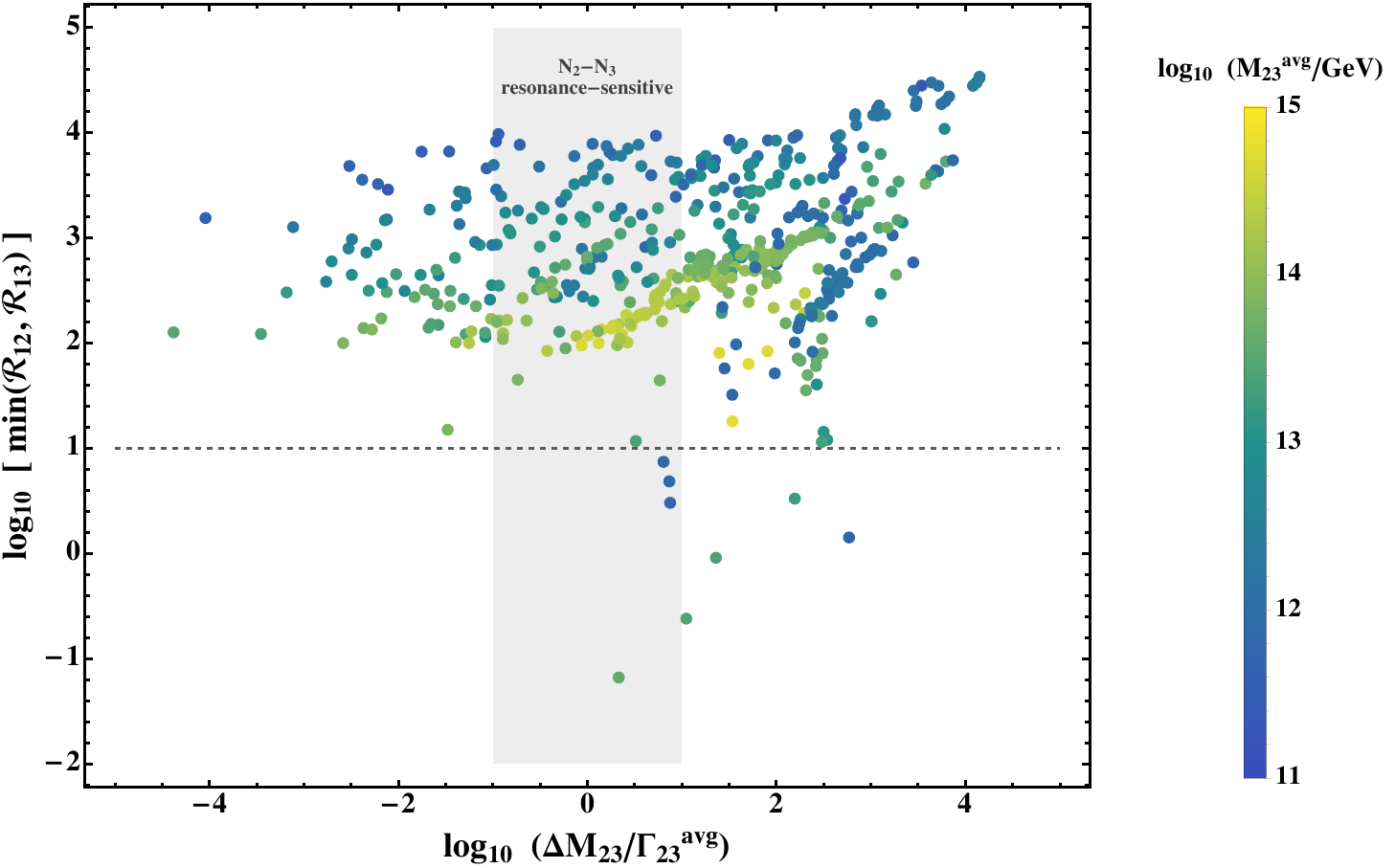}
\caption{Diagnostic of the quasi-degenerate RHN spectrum in the allowed parameter space. The horizontal axis shows $\log_{10}\mathcal{R}_{23}$, with the shaded region indicating the resonance sensitive range $0.1\leq\mathcal{R}_{23}\leq10$. The vertical axis shows $\log_{10}\mathcal{R}_{1(23)}$, where $\mathcal{R}_{1(23)}=\min(\mathcal{R}_{12},\mathcal{R}_{13})$, and therefore measures the separation of $N_1$ from the $(N_2,N_3)$ pair. The color scale denotes $\log_{10}(M_{23}^{\rm avg}/{\rm GeV})$, with $M_{23}^{\rm avg}=(M_{N_2}+M_{N_3})/2$. Points near $\mathcal{R}_{23}\sim1$ and with $\mathcal{R}_{1(23)}\gg1$ correspond to an isolated resonance-sensitive $(N_2,N_3)$ pair.}
\label{fig:quasidegenerate_diagnostic}
\end{figure}

This appendix collects the parameter-space diagnostics used in the leptogenesis analysis. Fig.~\ref{fig:RHNspectrum} shows the RHN mass spectra obtained from the viable normal-ordering parameter sets, while Fig.~\ref{fig:quasidegenerate_diagnostic} illustrates the quasi-degenerate $(N_2,N_3)$ region and its separation from $N_1$. These plots support the discussion of the allowed leptogenesis regimes in the main text.

%%%%%%%%%%%%%%%%%%%%%%%%%%%%%%%%%%%%%%%%%%%%%%%

%%%%%%%%%%%%%%%%%%%%%%%%%%%%%%%%%%%%%%%%%%%%%%%%
\FloatBarrier
\bibliographystyle{utcaps_mod}
\bibliography{A4_doubleV1}
\end{document}